\documentclass[fleqn,usenatbib]{mnras}

\usepackage{newtxtext,newtxmath}

\usepackage[T1]{fontenc}
\usepackage{ae,aecompl}
\usepackage{breqn}

\usepackage{graphicx}	
\usepackage{amsmath}	
\usepackage{amssymb}	
\usepackage{subfigure}
\usepackage{pdflscape}
\usepackage{caption}
\newcommand\rc{$\rm r_c$}
\newcommand\reff{$\rm R_{eff}$}
\newcommand\lya{Ly$\alpha$}
\newcommand\ha{H$\alpha$}
\newcommand\hb{H$\beta$}
\newcommand\sigmasfr{$\rm \Sigma_{SFR}$}

\newcommand\vssz{$v_s/\sigma_0$}
\graphicspath{{fig/}}
\title[Photometry of star-forming clumps in LARS galaxies]{Star-forming clumps in the Lyman Alpha Reference Sample of galaxies - I. Photometric analysis and clumpiness}

\author[M. Messa et al.]{Matteo Messa,$^{1}$\thanks{E-mail: matteo.messa@astro.su.se}
Angela Adamo,$^{1}$
G\"{o}ran \"{O}stlin,$^{1}$
Jens Melinder,$^{1}$
\newauthor
Matthew Hayes,$^{1}$
Johanna S. Bridge$^{2}$
and John Cannon$^{3}$
\\
$^{1}$Dep. of Astronomy, the Oskar Klein Centre, Stockholm University, Stockholm, Sweden\\
$^{2}$Dep. of Physics \& Astronomy, University of Luisville,  Dep. of Astronomy, Louisville, KY 40208, USA\\
$^{3}$Dep. of Physics \& Astronomy, Macalester College, St Paul, MN 55105, USA\\
}

\date{Accepted XXX. Received YYY; in original form ZZZ}

\pubyear{2019}

\begin{document}
\label{firstpage}
\pagerange{\pageref{firstpage}--\pageref{lastpage}}
\maketitle

\begin{abstract}
We study young star-forming clumps on physical scales of $10-500$ pc in the Lyman-Alpha Reference Sample (LARS), a collection of low-redshift ($z=0.03-0.2$) UV-selected star-forming galaxies.
In each of the 14 galaxies of the sample, we detect clumps for which we derive sizes and magnitudes in 5 $UV$-optical filters. The final sample includes $\sim1400$ clumps, of which $\sim600$ have magnitude uncertainties below 0.3 in all filters. 
The $UV$ luminosity function for the total sample of clumps is described by a power-law with slope $\alpha=-2.03^{+0.11}_{-0.13}$. Clumps in the LARS galaxies have on average \sigmasfr\ values higher than what observed in HII regions of local galaxies and comparable to typical SFR densities of clumps in $z=1-3$ galaxies. 
We derive the clumpiness as the relative contribution from clumps to the $UV$ emission of each galaxy, and study it as a function of galactic-scale properties, i.e. \sigmasfr\ and the ratio between rotational and dispersion velocities of the gas ($v_s/\sigma_0$). 
We find that in galaxies with higher \sigmasfr\ or lower $v_s/\sigma_0$, clumps dominate the $UV$ emission of their host systems.
All LARS galaxies with \lya\ escape fractions larger than $10\%$ have more than $50\%$ of the $UV$ luminosity from clumps. We tested the robustness of these results against the effect of different physical resolutions. At low resolution, the measured clumpiness appears more elevated than if we could resolve clumps down to single clusters. This effect is small in the redshift range covered by LARS, thus our results are not driven by the physical resolution.
\end{abstract}




\section{Introduction}
High-redshift ($\rm z\gtrsim1$) star-forming galaxies are morphologically dominated by clumpy structures \citep{cowie1995,vandenbergh1996} which can account for $\sim40\%$ of the galactic rest-frame $UV$ emission \citep[e.g.][]{elmegreen2005}. Such star-forming clumps have observed sizes $\sim0.1-1.5$ kpc and estimated masses $\rm M\sim10^8-10^9\ M_\odot$ \citep{elmegreen2007,guo2012,tacconi2013}. 
Their surface densities are $\sim8$ times the disk density and many of them form stars at high rate ($\rm SFR\sim0.5-100\ M_\odot/yr$, \citealt{forsterschreiber2011,genzel2011}). These characteristics make them very different to the `moderate' star-forming regions observed in the local universe. 
The difference must be set by the specific conditions of galaxies at high redshift: the majority of high-redshift galaxies show signs of rotation, indicating the presence of disks \citep{genzel2006,forsterschreiber2006,shapiro2008}, which have been observed to be highly turbulent \citep{cresci2009,forsterschreiber2009,wisnioski2015} and with gas-to-stellar mass ratios $>2$ times higher than in local disks \citep{tacconi2008,tacconi2013,saintonge2013}. The standard interpretation is that giant clumps are the result of gas collapse due to gravitational instabilities in the disk, which at high redshift can fragment at much larger scales because of its aforementioned properties \citep[e.g.][]{elmegreen2009,tamburello2015}. 

Recent studies on lensed galaxies have allowed the analysis of high-redshift galaxies down to $\sim100$ pc resolution \citep[e.g.][]{livermore2012,adamo2013,wuyts2014,cava2018} revealing that clumps are on average smaller and less massive ($\rm M\sim10^6-10^8\ M_\odot$) than previously thought. 
However, there are indications that clump properties evolve with redshift \citep{livermore2015}.

Massive clumps shape the morphologies of galaxies and possibly affect their evolution in the systems we observe locally. It is still not clear what are the time-scales for clump survival in their host galaxies. High-redshift clumps have usually estimated ages around $\sim100$ Myr, which can reach up to $\sim1$ Gyr \citep[e.g.][]{guo2012}. 
Models of clump evolution proposed that they can slowly migrate towards the centre of the galaxy, where they will eventually coalesce, contributing to the formation of the galactic bulge and of the thick disk \citep{bornaud2007,elmegreen2008}. Observations of single galaxies seem to support this scenario \citep[e.g.][]{guo2012,adamo2013,cava2018} but in general the test of such models is limited by the need for a thorough characterization of both clump properties and their position in the galaxy.

Feedback from young clumps can also affect the evolution of galaxies, as it is responsible for the suppression of global star formation and for the formation of a multi-phase interstellar medium (ISM) \citep[e.g.][]{hopkins2012,goldbaum2016}.
The impact of stellar feedback from clusters and clumps on galaxies can be so great that it may facilitate the escape of UV radiation into the inter-galactic medium \citep[e.g.][]{bik2015,bik2018}. For this reason, understanding the feedback process is fundamental to understand the escape of radiation from galaxies at high redshift and the reionisation of the universe \citep{bouwens2015}. 
Despite having such a big effect on the galaxy and its surroundings, the escape of UV radiation is possible only after clearing the dense clouds surrounding the very young star clusters \citep{dale2015,howard2018}. Both the escape of ionizing radiation and that of resonant lines (e.g. \lya) strongly depend on the gas distribution and conditions at sub-galactic scales. 

The study of local galaxies with properties resembling the ones at high redshift allows the exploration of physical scales impossible to resolve in more distant galaxies and therefore help the understanding of what regulates the star formation process.  
As an example, \citet{fisher2017} were able to study star-forming clumps on scales $\sim100$ pc in the DYNAMO sample \citep{green2014}, a collection of galaxies at $\rm z\sim0.1$ with high gas fractions ($\rm f_{gas}\sim20-40\%$, where $\rm f_{gas}\equiv M_{mol}/(M_{mol}+M_*)$, \citealt{fisher2014}) and \ha\ velocity dispersions $30-80$ km/s \citep{green2014,bassett2014}, similar
to those of high-redshift turbulent, clumpy disks. Due to the good characterisation of the clumps, the study of the DYNAMO sample was able to support the instability models describing star-formation at high redshift and in particular that the galaxy clumpiness is related to the ratio of velocity dispersion to rotation velocity \citep{fisher2017,fisher2017b}.  

The Lyman-Alpha Reference Sample (LARS) is a galaxy sample consisting of 14 low-redshift starburst systems ($z=0.03-0.2$) observed in multiple bands with the Hubble Space Telescope (HST) in order to study resolved \lya\ emission from low-redshift galaxies \citep{hayes2014,ostlin2014}. The galaxies are characterized in the $UV$ filters by star-forming clumps, which, due to the proximity of the galaxies, are resolved down to much smaller spatial scales ($\sim10$ pc) than the high-redshift systems. These observations can therefore help fill the gap between the star clusters studied in local galaxies (clusters have typical sizes $<10$ pc) and the more massive clumps observed in more distant galaxies (studied on scales $>100$ pc). The different morphological \citep{ostlin2014,guaita2015,micheva2018} and kinematic \citep{herenz2016} properties of the LARS galaxies can be used to test the clumpiness against the properties of the host galaxies. 
We divide the study of star-forming clumps in LARS galaxies in two parts: in this present work we study the photometric properties of the clumps, with the goal of understanding what affects the typical luminosities and surface brightnesses of clumps, as well as the clumpiness of the galaxies themselves. In a second work (Messa et al., in prep) we study physical properties (ages, masses, extinctions) of clumps, derived via broadband SED fitting, and study the ionizing budget of clumps, comparing it to the observations of \lya\ and \ha\ maps. This paper is divided as following: in Section~\ref{sec:2} we describe the sample selection, observation and some derived properties of the LARS sample, briefly summarizing previous works of the LARS collaboration. In Section~\ref{sec:photometry} we describe the extraction of the clump catalogue and its photometric analysis. In Section~\ref{sec:results} we present the results of the analysis. Finally, the main analyses and findings of this work are summarized in the conclusions, Section~\ref{sec:conclusions}.

\section{Sample of study and observations}
\label{sec:2}
The galaxies used in this study constitute the Lyman-Alpha Reference Sample (LARS), whose properties and selection are extensively described in \citet{ostlin2014}. We report here a summary of the sample selection, observations and main galaxy properties.

\subsection{Sample selection}
\label{sec:samplesel}
LARS is a sample of 14  galaxies in the low-redshift universe, with redshifts spanning the range $\rm z=0.028-0.18$. They were selected from a cross-match between SDSS(DR6) and $GALEX$(DR3) catalogues as star-forming galaxies, with EW(H$\alpha)>100$ \AA. It was found that this criterion leads to samples dominated by compact systems,
mainly of irregular morphology \citep{heckman2005}, whereas lowering the EW(H$\alpha$) limit would favour the inclusion of more ordinary-looking disk galaxies. Galaxies with strong active galactic nuclei (AGN) were rejected by selecting galaxies with narrow \ha\ line widths (line-of-sight $\rm FWHM<300$ km/s) and based on their position on the BPT diagram \citep{baldwin1981}. In selecting the targets following these two main criteria, an effort was made to cover a wide range of FUV luminosities and priority was given to low-z galaxies (in order to better characterize sub-galactic scales) and to galaxies with HST archival data.
The selected sample of 14 galaxies is listed in Tab~\ref{tab:lars}, together with their main properties derived by SDSS and $GALEX$.
They span a FUV luminosity range between $\log(\nu \textrm{L}_\nu/\textrm{L}_\odot)=9.2$ and $\log(\nu \textrm{L}_\nu/\textrm{L}_\odot)=10.7$, which  encompasses that of high-z Lyman-$\alpha$ emitters \citep[LAEs, e.g.][]{nilsson2009}, $GALEX-$selected LAEs \citep{deharveng2008}, and
$\rm z \sim 0.3$ $GALEX-$selected Lyman-break galaxy (LBG) analogues \citep{hoopes2007,overzier2008}.
\begin{landscape}
\begin{table}
\centering
\caption{Properties of the LARS galaxies, from \citet{ostlin2014} (columns 1-5), this work (columns 6-9), \citet{herenz2016} (column 10-11) and Melinder et al., in prep. (column 12). The values in the columns are: (1) redshift; (2) \ha\ equivalent width; (3) $UV$ luminosity; (4) oxygen abundance in units of 12+log(O/H), determined utilizing the temperature-sensitive $\rm [O III]_{4363}$ line; (5) oxygen abundance derived by the empirical O3N2 relation;
(6) galactic radius ($r_g\equiv\sqrt[]{A/\pi}$); (7) SFR derived from $FUV$ luminosity; (8) stellar mass of the galaxy, summing the contribution of old and young stellar populations; (9) average SFR surface density (derived from values in column 4 and 5); 
(10) dispersion velocities of ionized gas (derived from \ha\ emission); 
(11) ratio between shear and dispersion velocities of ionized gas (derived from \ha\ emission); (12) escape fraction of \lya\ radiation. More details are given in the text.}
\label{tab:lars}
\begin{tabular}{rrcrrrrrrrrrr}
\hline
\multicolumn{1}{c}{ID} & \multicolumn{1}{c}{z} & \multicolumn{1}{c}{W(\ha)} & \multicolumn{1}{c}{$\rm \log(L_{UV}/L_\odot)$} &
\multicolumn{1}{c}{O/H$\rm _{T_e}$} & \multicolumn{1}{c}{O/H$\rm _{O3N2}$} & \multicolumn{1}{c}{$\rm r_g$} & \multicolumn{1}{c}{$\rm SFR_{UV}$} & 
\multicolumn{1}{c}{$\rm M_*$ } & \multicolumn{1}{c}{$\rm \Sigma_{SFR}$} & \multicolumn{1}{c}{$\sigma_0$} & \multicolumn{1}{c}{$\textrm{v}_{\textrm{shear}}/\sigma_0$} & \multicolumn{1}{c}{$\textrm{f}_{\textrm{esc}}(\textrm{Ly}\alpha)$} \\
\multicolumn{1}{c}{\ } & \multicolumn{1}{c}{\ } & \multicolumn{1}{c}{(\AA)} & \multicolumn{1}{c}{\ } &
\multicolumn{1}{c}{\ } & \multicolumn{1}{c}{\ } & 
\multicolumn{1}{c}{(kpc)} & \multicolumn{1}{c}{($\rm M_\odot/yr$)} & \multicolumn{1}{c}{($\rm 10^{10}\ M_\odot$)} & 
\multicolumn{1}{c}{($\rm M_\odot/yr/kpc^2$)} & \multicolumn{1}{c}{(km/s)} & \multicolumn{1}{c}{\ } & \multicolumn{1}{c}{\ } \\
\multicolumn{1}{c}{} & \multicolumn{1}{c}{(1)} & \multicolumn{1}{c}{(2)} & \multicolumn{1}{c}{(3)} & 
\multicolumn{1}{c}{(4)} & \multicolumn{1}{c}{(5)} & \multicolumn{1}{c}{(6)} & \multicolumn{1}{c}{(7)} & 
\multicolumn{1}{c}{(8)} & \multicolumn{1}{c}{(9)} & \multicolumn{1}{c}{(10)} & \multicolumn{1}{c}{(11)} & \multicolumn{1}{c}{(12)} \\
\hline
01 & 0.028 & 575 & 9.92 	& $8.070$  & $8.242$   & $2.70$   &  $7.45\ \pm0.16$    &  $1.096\ \pm0.014$	& $0.3253\ \pm0.0071$	& $47.5\ \pm0.1$ & $1.2\ \pm0.1$		& $0.205\ \pm0.003$ \\
02 & 0.030 & 315 & 9.48 	& $8.041$  & $8.226$   & $3.60$   &  $1.43\ \pm0.09$    &  $0.560\ \pm0.012$	& $0.0351\ \pm0.0022$	& $38.6\ \pm0.9$ & $0.6\ \pm0.1$		& $0.444\ \pm0.011$ \\
03 & 0.031 & 241 & 9.52 	& $...$  & $8.414$   & $6.97$   & $23.25\ \pm1.29$    &  $2.694\ \pm0.268$	& $0.1523\ \pm0.0084$	& $99.5\ \pm3.7$ & $1.4\ \pm0.2$		& $0.008\ \pm0.001$ \\
04 & 0.033 & 237 & 9.93 	& $8.191$  & $8.191$   & $5.86$   &  $2.74\ \pm0.01$    &  $2.477\ \pm0.013$	& $0.0254\ \pm0.0001$	& $44.1\ \pm0.1$ & $1.7\ \pm0.1$		& $0.000\ \pm0.002$ \\
05 & 0.034 & 340 & 10.01   	& $7.800$  & $8.124$   & $2.17$   &  $3.40\ \pm0.01$    &  $0.862\ \pm0.003$	& $0.2299\ \pm0.0008$	& $46.8\ \pm0.3$ & $0.8\ \pm0.1$		& $0.174\ \pm0.004$ \\
06 & 0.034 & 464 & 9.20		& $7.864$  & $8.082$   & $4.96$   &  $0.47\ \pm0.02$    &  $0.474\ \pm0.011$	& $0.0061\ \pm0.0002$	& $27.2\ \pm0.3$ & $1.9\ \pm0.2$		& $0.000\ \pm0.012$ \\
07 & 0.038 & 434 & 9.75 	& $7.911$  & $8.352$   & $2.28$   &  $4.02\ \pm0.10$	 &  $0.619\ \pm0.006$	& $0.2462\ \pm0.0058$	& $58.7\ \pm0.3$ & $0.5\ \pm0.1$		& $0.156\ \pm0.004$ \\
08 & 0.038 & 170 & 10.15   	& $...$  & $8.505$   & $7.63$   & $39.01\ \pm5.09$  	 &  $8.095\ \pm0.280$	& $0.2133\ \pm0.0278$	& $49.0\ \pm0.1$ & $3.2\ \pm0.1$		& $0.008\ \pm0.001$ \\
09 & 0.047 & 522 & 10.46	& $8.051$  & $8.366$   & $7.17$   & $31.12\ \pm0.68$  	 &  $5.199\ \pm0.039$	& $0.1927\ \pm0.0042$	& $58.6\ \pm0.1$ & $3.1\ \pm0.1$		& $0.025\ \pm0.001$ \\
10 & 0.057 & 100 & 9.74		& $...$  & $8.505$   & $5.49$   &  $2.42\ \pm0.13$    &  $2.130\ \pm0.030$	& $0.0256\ \pm0.0014$	& $38.2\ \pm1.0$ & $0.9\ \pm0.2$		& $0.020\ \pm0.004$ \\
11 & 0.084 & 108 & 10.70   	& $...$  & $8.436$   & $8.93$   & $23.23\ \pm1.16$  	 & $12.436\ \pm0.176$	& $0.0927\ \pm0.0046$	& $69.3\ \pm3.8$ & $2.1\ \pm0.3$		& $0.085\ \pm0.004$ \\
12 & 0.102 & 447 & 10.53   	& $8.008$  & $8.342$   & $2.71$   & $13.50\ \pm0.24$    &  $2.788\ \pm0.030$	& $0.5850\ \pm0.0106$	& $72.7\ \pm1.0$ & $1.3\ \pm0.1$		& $0.041\ \pm0.002$ \\
13 & 0.147 & 226 & 10.60   	& $...$  & $8.503$   & $5.15$   & $15.24\ \pm2.19$  	 &  $3.904\ \pm0.218$	& $0.1829\ \pm0.0263$	& $69.2\ \pm0.7$ & $2.5\ \pm0.2$		& $0.010\ \pm0.008$ \\
14 & 0.181 & 605 & 10.69   	& $7.823$  & $8.055$   & $1.89$   & $13.92\ \pm0.12$  	 &  $0.711\ \pm0.014$	& $1.2404\ \pm0.0104$	& $67.3\ \pm1.3$ & $0.6\ \pm0.1$		& $0.358\ \pm0.017$ \\
\hline
\end{tabular}
\end{table}
\end{landscape}

\subsection{HST observations}
LARS galaxies were imaged using multi-band observations with HST. The filters were chosen to be able to characterize emission from stellar continuum as well as from \ha, \hb\ and \lya\ lines. \ha\ and \hb\ were observed via narrowband filters. Emission from \lya\ was imaged using a synthetic filter made from the combination of long-pass filters of the solar-blind channel (SBC) camera, F125LP and F140LP for LARS01 to LARS12 and F140LP and F150LP for LARS13 and LARS14. The stellar continuum was sampled near the \lya\ line with the F150LP filter and on both sides of the 4000 \AA\ break in $U$ and $B$ bands \citep{hayes2009}.
This filter combination allows us to derive ages and masses for the stellar component of the galaxy via broadband SED fitting. In order to obtain a better accuracy in the SED fit, a filter at redder wavelengths is also included ($i$ band). A typical filter set for LARS observations consist  therefore of ACS\footnote{ACS: Advanced Camera for Surveys.}/SBC F125LP, F140LP and F150LP, WFC3\footnote{WFC3: Wide Field Camera 3.}/UVIS F336W, F438W, F775W, F502N and F673N. This is, for example, the set of filters used for observing all LARS galaxies between LARS01 and LARS12, excluding LARS03. Even though no conversion is applied to the Johnsons-Cousins filter system, for the reminder of the paper we keep the same nomenclature, due to the similarity of the central wavelength between that system and our data, in particular referring to F336W, F438W, F775W filters as $U$, $B$ and $i$ band respectively.
A complete list of filter sets and exposure times for the LARS observations is given in Tab.~4 of \cite{ostlin2014}.
LARS13 and LARS14 count one filter less than the other galaxies, since observations in the F125LP filter were not necessary. The LARS galaxies as observed in the F140LP filter (corresponding to rest-frame wavelength $\sim1500$\AA) are shown in Fig.~\ref{fig:lars_f140lp}. All data were drizzled to the same pixel scale of $0.04$ arcsec/pixel.
\begin{figure*}
\centering
\includegraphics[width=1.\textwidth]{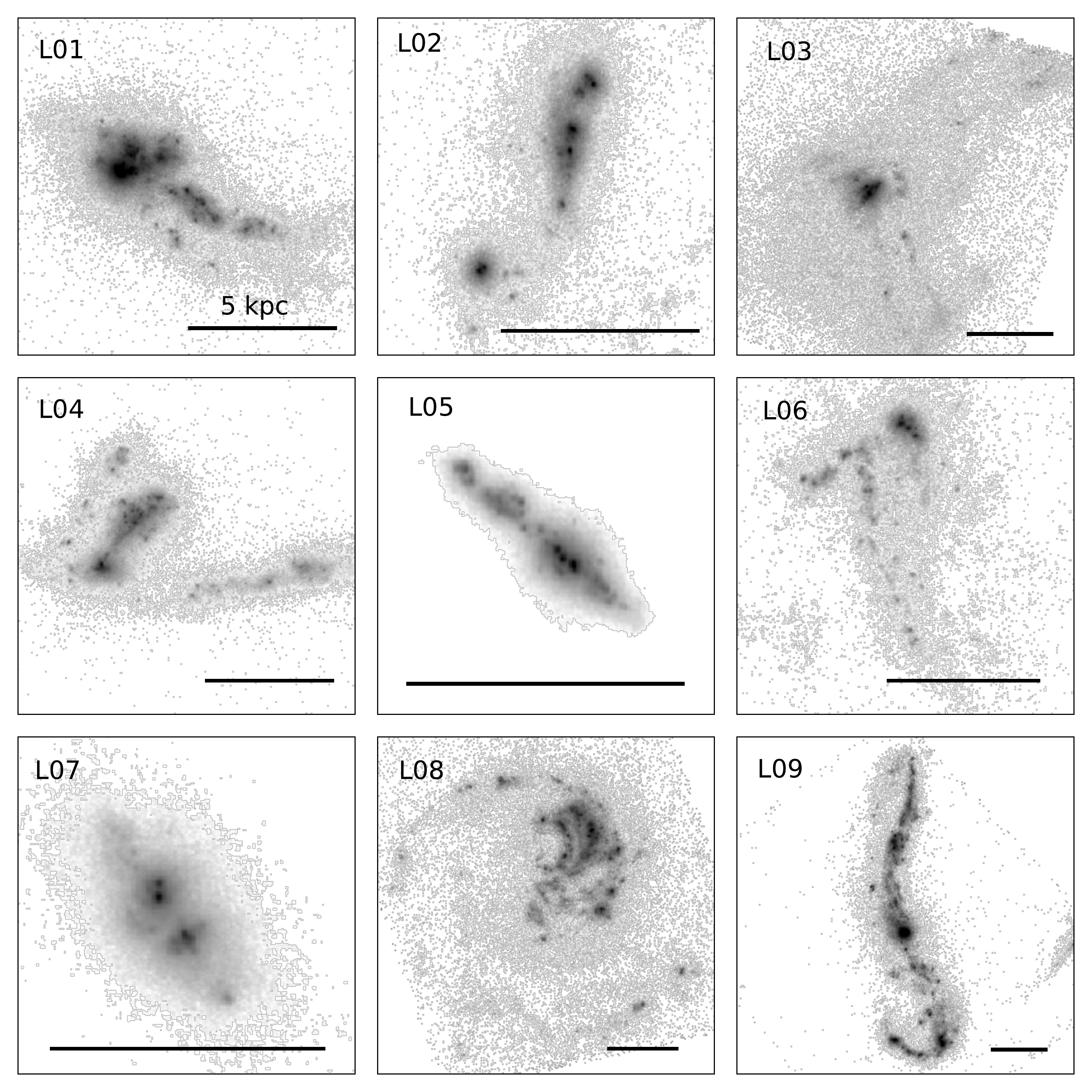}
\caption{Zoom-in of the LARS sample as observed in the F140LP filter (F150LP for LARS13 and LARS14), corresponding to a rest-frame wavelength of $\sim1500$ \AA. The $UV$-bright star forming clumps are visible in all galaxies. North is up and east to the left. Black bars in
the lower right corner show a physical size of 5 kpc.}
\label{fig:lars_f140lp}
\end{figure*}
\begin{figure*}\ContinuedFloat
\centering
\includegraphics[width=\textwidth]{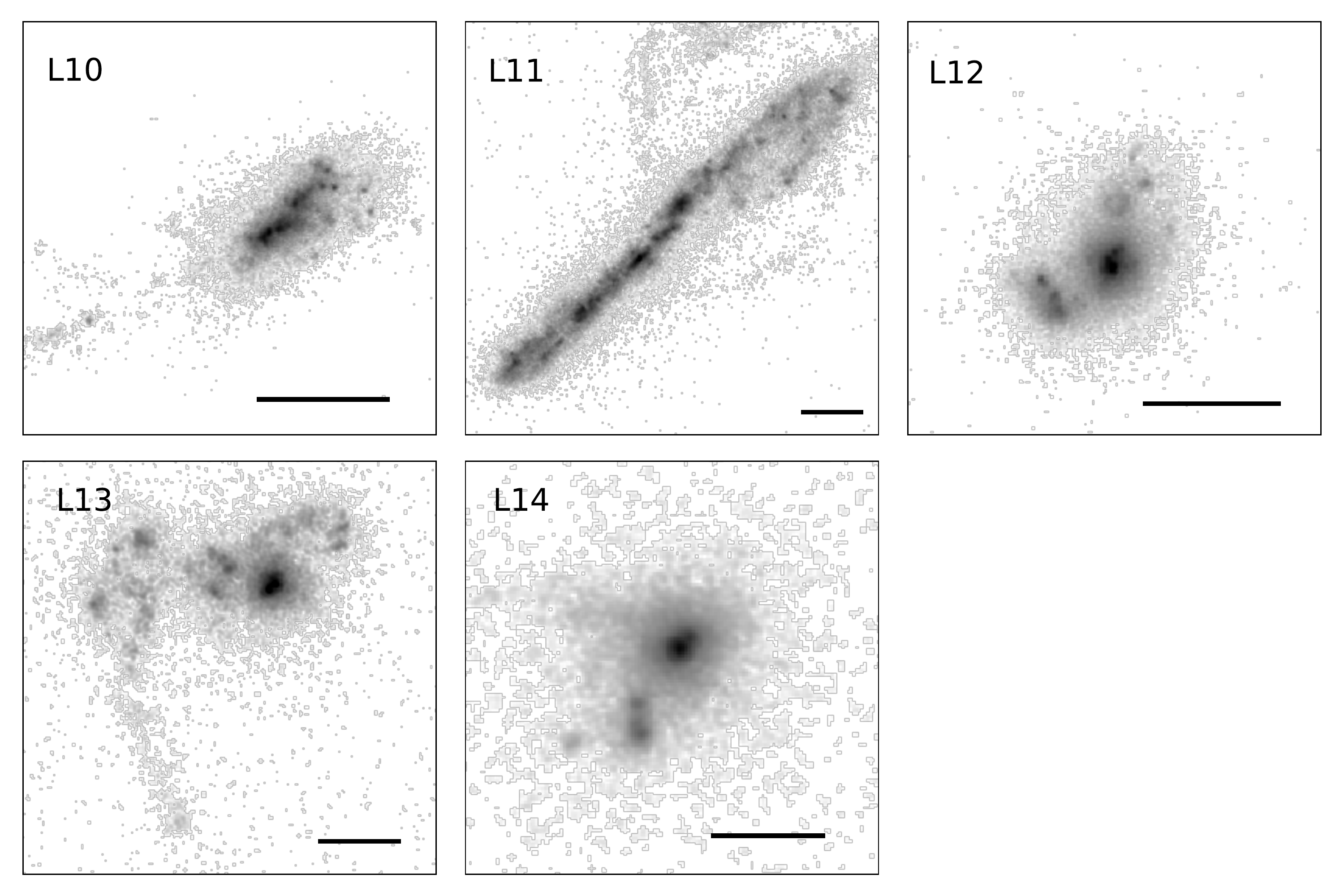}
\caption{Continued.}
\end{figure*}

\subsection{Ancillary observations}
In order to characterize the ISM inside and around the LARS galaxies, the sample was observed in different wavelength ranges with other facilities, such as the 100-m Green Bank Telescope (GBT) and the Karl G. Jansky Very Large Array (VLA) to spectroscopically study the HI emission from the neutral gas \citep{pardy2014} and the Potsdam Multi-Aperture Spectrophotometer (PMAS) to study resolved \ha\ kinematics \citep{herenz2016}. The HI observations available are on much larger scales ($\sim10$ times) than the size of the galaxies themselves and are therefore characterizing the properties of the neutral gas surrounding the galaxies. On the other hand, PMAS observation have spatial resolution around $\sim 1''$ (equivalent to $\sim0.6$ kpc in the closest galaxy and to $\sim3$ kpc in the furthest one) and allowed a sub-galactic analysis of LARS galaxies. Details of the observations, data reduction and data analysis of the PMAS data are given in \citet{herenz2016} and we report here only the derived global properties of the \ha\ kinematics, in order to use them to characterize the galaxies in the analyses of this current work. Values of these global properties are listed in Tab.~\ref{tab:lars}. In more detail,
the shearing velocity, $v_s$, which measures large-scale gas bulk motions along the line of sight, was calculated considering the minimum and maximum velocities of \ha, via $\rm v_s = (v_{max}-v_{min})/2$, without inclination corrections. The second global property derived is the intrinsic velocity dispersion, $\sigma_0$, which measures the strength of random motions of the ionized gas. It is derived as the flux-weighted average of the velocity dispersion in each spaxel in the galaxy. Combining these two parameters we can use the ratio $v_s/\sigma_0$ as a quantification of whether the gas kinematic is dominated by ordered or turbulent motions. Usually galaxies with $v_s/\sigma_0<1$ are considered dispersion-dominated systems. We point out that high values of $v_s$ do not necessarily imply the presence of rotating disks.

\subsection{Summary of the LARS derived properties}
\label{sec:lars_properties}
The main goal of the LARS project is the study of geometrically resolved \lya\ emission from star-forming galaxies. To accomplish this, the broad and narrow-band observations are used to fit the stellar and nebular content of the galaxy. Once the contribution of the stellar continuum in the bluest FUV filter is derived and subtracted, the observation in that same filter is used to image the \lya\ emission of the galaxy. 

In order to do so, the frames in various filter are reduced, aligned and PSF-matched (i.e. all are convolved in order to have the same synthetic PSF). The method is explained in detail in Melinder et al. (in prep), where an updated analysis of the LARS galaxies, together with the extension of the project (eLARS) consisting of 28 additional galaxies, will be given.
As by-products of the SED fitting we produce maps of some important properties of the galaxies, such as \ha\ (continuum subtracted) maps and stellar masses. An overview of these properties for the LARS and eLARS samples are given in \citet{hayes2014} and Melinder et al. (in prep). We describe here some of the properties derived in the analysis of Melinder et al. (in prep), which will be used in this work. 

First of all, since we want to consider global properties, we need to define the physical extent of each galaxy. The extension of the galaxy was  measured on the FUV continuum images, binned using a Voronoi tessellation algorithm in order to maintain sufficient signal-to-noise in the outskirts. We define the area over which we measure global properties by including all spatial bins with fluxes greater than the isophotal level that encompass $90\%$ of the FUV continuum flux. In this way the clumps are included in the considered region while high-flux noise peaks outside the galaxies are neglected. 
The star formation rate (SFR) of the galaxy is derived by summing the FUV continuum flux (extinction corrected) emission inside the galaxy area and using the \citet{kennicutt2012} conversion. Stellar mass maps comes from the SED fitting of the Voronoi tessellated maps of the galaxies, where two simple stellar populations (an old one and a young one) are modelled. The mass of both are summed inside the galaxy area. Radii, SFRs and stellar masses, $\rm M_*$ are listed in Tab.~\ref{tab:lars}. The radius reported on the table correspond to one of a circularized region with the same surface as the area of the galaxy we are considering ($r_g\equiv\sqrt[]{A/\pi}$). We list in the last column of Tab.~\ref{tab:lars} the values for the escape fractions of \lya\ radiation, $\rm f_{esc}$(\lya). In order to take into account the scattering of \lya\ away from the galaxy, this is measured in a larger region (optimized to get the highest possible signal-to-noise in \lya) than the one used for SFR and $\rm M_*$. A detailed description about the calculation of $\rm f_{esc}$(\lya) is given in Melinder et al. (in prep). Other properties of the LARS galaxies, mainly derived by SDSS data, can be found in \citet{ostlin2014}.

\subsection{Comparison to other low and high-redshift samples}
\label{sec:samples_comparison}
The LARS galaxies were put in the broader context of other galaxy populations in \citet{ostlin2014} and in \citet{guaita2015}. Reporting the considerations discussed in those works, LARS galaxies can be studied as reference of $\rm 10^9-10^{11}\ M_\odot$ high-z (2<z<3)
star-forming galaxies (e.g. \citealt{law2012}). They also have continuum sizes and stellar masses similar to those of local starburst Lyman break analogues at $z\sim0.2$ (e.g. \citealp{overzier2009,overzier2010}). 
In this current work we will compare the stellar clumps of the LARS galaxies to other samples at different redshift. In particular, another local sample of galaxies used as reference for high-redshift systems is the DYNAMO sample \citep{green2014}, where stellar clumps have been studied down to physical scales of $\sim30$ pc \citep{fisher2017}. Differently from our sample, the DYNAMO galaxies used in the \citet{fisher2017} study have been selected to host disks with significant velocity gradients and large gas velocity dispersions in the outer disk, $\sigma_{H\alpha}\sim 30-80$ km/s, resulting in a sample dominated by strong rotators. On the other hand, LARS galaxies were selected on $H\alpha$ equivalent width and FUV emission, in order to ensure galaxies with recent bursts of star formation, but without imposing any constraint on the galaxy dynamics. The resulting sample is dominated, in the morphology and in the dynamics, by irregular and merging systems \citep{guaita2015,herenz2016}. Because of their irregular morphologies and elevated star formation, LARS galaxies could be analogs of merging systems, like the Antennae, which have been observed to host massive star clusters and cluster complexes (e.g. \citealp{whitmore1999,wilson2000}).

\section{Clump extraction and photometry}
\label{sec:photometry}
We created a catalogue of clumps by running the \texttt{SExtractor} software \citep{sextractor} on each of the galaxies. The $B$-band filter was used as the reference filter for the extraction, as it constitutes a good compromise between the UV, where the young clumps are brightest but the extinction is also the highest, and the longer wavelengths, where the extinction is lower but the clumps are mixed with the older stellar field of the galaxies, which dominates the light at these wavelengths.

When finding the sources with \texttt{SExtractor} we noticed two critical aspects that strongly affect the process. First, LARS galaxies have diffuse light emission of very different surface brightnesses, often caused by their disturbed morphologies, and this requires the ability of extracting clumps located at different background levels. In some cases, the background level changes on scales of a few pixels.  
In addition, clumps are often clustered together, with the necessity of high deblending factors. For these reasons, in order to avoid missing sources, for half of the galaxies we ran \texttt{SExtractor} twice, with different input parameters and in particular with different sizes of the rectangular grid inside which local background is estimated (Ba). Other key parameters of the extraction were the threshold signal-to-noise per pixel which defined a detection ($\rm D_T$), the deblending parameter for nearby sources (Deb, i.e. the minimum flux fraction that the branch of a composite object should have in order to be considered as a separate object) and minimum number of pixels above the threshold ($\rm D_M$), which was kept fixed at 5 for all the runs. The choice of doing a second run was based on a visual inspection of the extracted sources.
Details of the configuration files used in the extraction are given in Tab.~\ref{tab:sextractor}. To give an example of how the extraction is sensitive to the these parameters, in LARS01 changing the signal-to-noise requested for a detection from $\rm D_T=3$ to $\rm D_T=4$ reduces by almost half the number of extracted sources. Changing the background size has a smaller effect, as, on the same galaxy, doubling background size reduces by $\sim20\%$ the number of extracted sources, while the deblending parameter has even a smaller effect, affecting only less than $5\%$ of the sources.
The extracted catalogue was visually inspected to remove clear interlopers such as bright pixels at the border of the chip and clumps belonging to nearby galaxies in the HST frame (in LARS09 and LARS11\footnote{LARS09 and LARS11 have companion galaxies, mainly falling outside the field of view of the SBC.}). \\
\begin{table}
\centering
\caption{Parameters used for the source extraction with \texttt{SExtractor}. In some galaxies two runs of the code were needed. The parameter listed in the table are $\rm D_M$:DETECT\_MINAREA (minimum number of pixels above
threshold triggering detection), $\rm D_T$:DETECT\_THRESH (threshold S/N), Deb:DEBLEND_MINCONT (Minimum contrast parameter for deblending), Ba:BACK_SIZE (size in pixels of a background mesh). Dashes mean that the second run was not necessary.} 
\label{tab:sextractor}
\begin{tabular}{lllllllll} %
\hline
Gal. 	& \multicolumn{4}{l}{Run 1} & \multicolumn{4}{l}{Run 2} \\
\ 		& $\rm D_M$	& $\rm D_T$	& $\rm Deb$	& $\rm Ba$	& $\rm D_M$	& $\rm D_T$	& $\rm Deb$	& $\rm Ba$ \\
\hline
L01 	& 5	& 3		& 0.005		& 5 	& $-$	& $-$	& $-$	& $-$ 		\\
L02 	& 5	& 3		& 0.005		& 5 	& $-$	& $-$	& $-$	& $-$ 		\\
L03 	& 5	& 4		& 0.001		& 5 	& 5		& 4		& 0.001	&20 		\\
L04 	& 5	& 4		& 0.001		& 5 	& 5		& 4		& 0.001	&20 		\\
L05 	& 5	& 4		& 0.001		& 5 	& 5		& 4		& 0.001	&10 		\\
L06 	& 5	& 2.5	& 0.001		& 20	& 5		& 2.5	& 0.001	&500 		\\
L07 	& 5	& 4		& 0.0001	& 5 	& $-$	& $-$	& $-$	&$-$ 		\\
L08 	& 5	& 4		& 0.0001	& 5 	& $-$	& $-$	& $-$	&$-$ 		\\
L09 	& 5	& 4		& 0.001		& 5 	& $-$	& $-$	& $-$	&$-$ 		\\
L10 	& 5	& 4		& 0.001		& 5 	& 5		& 4		& 0.001	&20 		\\
L11 	& 5	& 4		& 0.001		& 5 	& 5		& 4		& 0.001	&20 		\\
L12 	& 5	& 3		& 0.001		& 5 	& $-$	& $-$	& $-$	&$-$ 		\\
L13 	& 5	& 3		& 0.001		& 5 	& 5		& 3		& 0.001	&10 		\\
L14 	& 5	& 3		& 0.005		& 5	    & $-$	& $-$	& $-$	&$-$		\\
\hline
\end{tabular}
\end{table}

On each of the extracted sources we performed a photometric analysis on filters F140LP, F150LP and on $U$, $B$ and $I$ bands (in LARS 13 and 14 the filter F140LP is used for observing the \lya\ emission and is therefore neglected in the clump photometric analysis). 
We report here the main steps of the photometric analysis (see Appendix~\ref{sec:a1} for further details and the description of a series of tests on completeness and uncertainties).
Each source was analysed in a $7\times7$ pixel cut-out, large enough to contain most of its flux but avoiding too strong contamination from neighbouring sources. Smaller boxes were tested, but contain too little signal to account for all the free parameters considered in the modelling. The clumps were modelled with a circular EFF-Moffat profile \citep{elson1987} of index 1.5 (kept fixed for all sources) and effective radius $\rm R_{eff}$. The Moffat profile was shown to be the best fit to the light profile of young massive clusters \citep{elson1987,bastian2013} because of its property of having broad emission wings. Ellipticity was not taken into account as many clumps in LARS galaxies are already well-fitted by circularly symmetric profiles, and it would require two additional free parameters. The Moffat profile was convolved with the instrumental PSF (K) to create a model of the \textit{observable} source. We point out once more that the PSFs of different filters were matched and therefore K is the same for all filters. In order to account for the presence of diffuse background emission (possibly coming from the diffuse stellar population, neighbouring sources or nebular emission), we added a $\rm 1^{st}$ degree polynomial background (described by the parameters $\rm c_0,\ c_x\ and\ c_y$). The \textit{observable} model (M) is therefore parametrized as:
\begin{multline}
M(x,y|x_0,y_0,F,r_c,c_0,c_x,c_y) = \\ \left[K\ast\left(F/F_0\cdot\left(1+(r/r_c)^2\right)^{-1.5}\right)\right]+c_0+c_xx+c_yy
\end{multline}
where \rc, called \textit{core} radius is related to the effective radius \reff\ by $\rm R_{eff}=\sqrt{3}r_c$\footnote{The effective radius ($\rm R_{eff}$), or \textit{half-light} radius, is the radius enclosing half of the source's flux.}, and the radial distance $r$ is defined as $r=\sqrt{(x-x_0)^2+(y-y_0)^2}$, where $x,y$ are the pixel coordinates and $x_0,y_0$ are the source centre coordinates inside the box. $F$ and \rc\ parametrize the flux and the size of the source, respectively (the Moffat profile is normalized via the factor $F_0$). 

\begin{figure}
\centering
\includegraphics[width=1.\columnwidth]{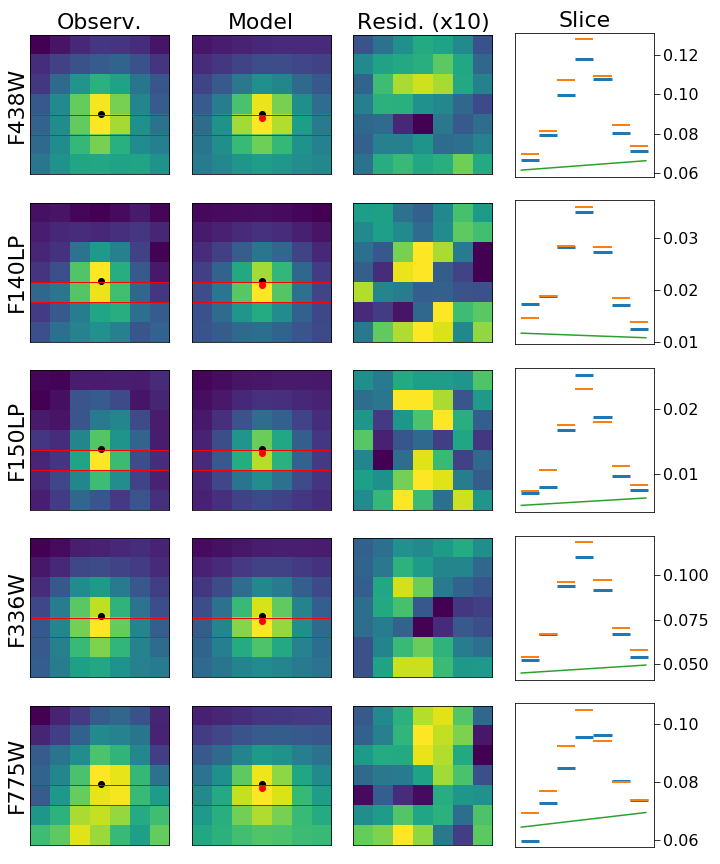}
\caption{Visual example of the photometric fit of a source (in LARS01): observations in the 5 bands ($\rm 1^{st}$ col), best models ($\rm 2^{nd}$ col) and residuals ($\rm 3^{rd}$ col). The $\rm 4^{th}$ column shows the fluxes of the observation (blue), model (orange) and background only (green) for each pixel of the highlighted horizontal slice. Fluxes for observation and model are plotted with the same colour scale, while the residuals fluxes were rescaled of a factor 10.}
    \label{fig:phot}
\end{figure}

We assume each clump to have the same size in all the bands and we fit the model to the cut-out image (over 7x7 pixels) in the 5 bands simultaneously (4 bands for LARS 13 and 14), keeping the same core radius \rc\ (and same centre coordinates $x_0,y_0$) but allowing the other parameters to vary from filter to filter. 
The best-fit values for the parameters were found by minimizing the residuals of the difference between the observable models and the data. We built a probability function based on the sum of the weighted residuals	
\begin{equation}
\ln{P} = -0.5\sum_{f,i}\left[\left(M_{f,i}-D_{f,i}\right)^2\cdot w_{f,i}\right]
\end{equation}
where $D_{f,i}$ are the data, $w_{f,i}=1/\sigma_{f,i}^2$ the weights and the sum is done for all pixels $i$ in the five bands $f$.
The probability function is used to run a Markov chain Monte Carlo (MCMC) sampling. We used the \texttt{Python} package \texttt{emcee} \citep{emcee}, which implements the MCMC sampler from \citet{goodman2010}, and run 50 walkers, each producing 320-step chains but discarding the first 20 steps from each, for a total of 15000 sampling values for each source. We consider the median value of the distribution of each parameter as its best-fit value. Uncertainties on each parameter were found to be symmetric and we consider half the difference between the 15.8 and 84.2 percentiles of each parameter distribution as its uncertainty. 
Each walker is an independent series and its starting value was chosen drawing it from a normal distribution centred on the maximum-likelihood value, calculated via a \textit{least-square} fit using the \texttt{Python} package \textit{lmfit}, and with a standard deviation equal to $20\%$ of the maximum-likelihood value.
Data, best fit models and residuals for an example clump are presented in Fig.~\ref{fig:phot} . 

Due to the limited size of the box used for photometry and to the resolution of our data, we were not able to derive the size of some of the sources, and in particular:
\begin{enumerate}
\item Our fitting routine is not sensitive to \rc\ values smaller than 0.3 pixels (see Appendix~\ref{sec:lowersize}). Sources whose recovered \rc\ is smaller than this value are assumed to have $\rm r_c=0.3$ as an upper limit.
\item For sources with \rc\ larger than 3.5 pixels, the \textit{core} region of the clump is not entirely enclosed in the fitting box. We could not trust the fit of those sources, which therefore were removed from the catalogue. This cut only affects 117 sources out of 1698 ($6.9\%$ of the total) and, by visual inspection, these sources turn out to be mainly artefacts of the diffuse emission.
\end{enumerate}

We convert the best-fit fluxes ($\rm F_{fit}$) in each filter into AB magnitudes ($\rm m_{AB}$) and the flux uncertainties $\rm F_{err}$ into magnitude uncertainties ($\rm m_{err}$) via $\rm m_{err}=1.0857\times F_{err}/F_{fit}$.
Finally, not all the sources were detected in all bands. We discarded from the sample all the sources with observed magnitude $\rm m_{AB}>30$ mag in more than 1 band. This cut reduces the total clump catalogue to 1425 sources, which can have magnitude uncertainties up to $\rm m_{err}\sim0.6$ mag in some filters. In the analyses of the following sections we also select a sub-sample of sources which are well-detected in all bands. We consider in this high-fidelity ($HF$) sample only clumps with $\rm m_{err}<0.3$ mag in all 5 bands (4 bands for LARS 13 and 14). The $HF$ sample includes a total of 608 clumps.
This further constraint implies that clumps in the $HF$ catalogue have magnitudes $\lesssim 26$ mag.

\subsection{Using ESO 338-IG04 as test bench for different resolutions}
In order to assess how loss in spatial resolution within our galaxy redshift range affects our analyses,
we consider a nearby galaxy observed with a set of filters similar to the LARS sample and degrade its resolution to simulate its observation at the redshifts of the LARS galaxies. We use the nearby starburst dwarf ESO 338-IG04 (also known as Tololo 1924-416, and hereafter shortened to ESO\,338), which hosts a population of young star clusters, with masses up to $\rm 10^7\ M_\odot$ \citep{ostlin1998,ostlin2003,adamo2011}. ESO 338 is at a measured distance of 37.5 Mpc \citep{ostlin1998} and was observed with HST in filters F218W, F336W, F439W and F814W of the Wide Field and Planetary Camera 2 \citep{ostlin1998} and with the ACS camera \citep{ostlin2009} in filters 
F122M and F140LP of the SBC (program GO 9470, PI: D.Kunth) and in filters F550M and FR565N of the WFC camera (program GO 10575, PI: G.\"Ostlin). As for the LARS galaxies, all the frames of ESO 338 were reduced, aligned and convolved to have the same hybrid PSF in all filters.  
The pixel scale for all the frames is $0.04''$, which at the distance of ESO 338 corresponds to $7.3$ pc.

We simulate ESO 338 observations at the distance of LARS01 and LARS14 (the closest and the most distant galaxies of our sample, respectively). LARS01 is at a distance of $119.7$ Mpc and has a PSF with full width at half maximum $\rm FWHM=2.6$ px, or $\rm FWHM\approx60$ pc. On the other hand the PSF of LARS14 has a $\rm FWHM\approx400$ pc (3.2 px). 
\begin{figure*}
\centering
\includegraphics[width=1.\textwidth]{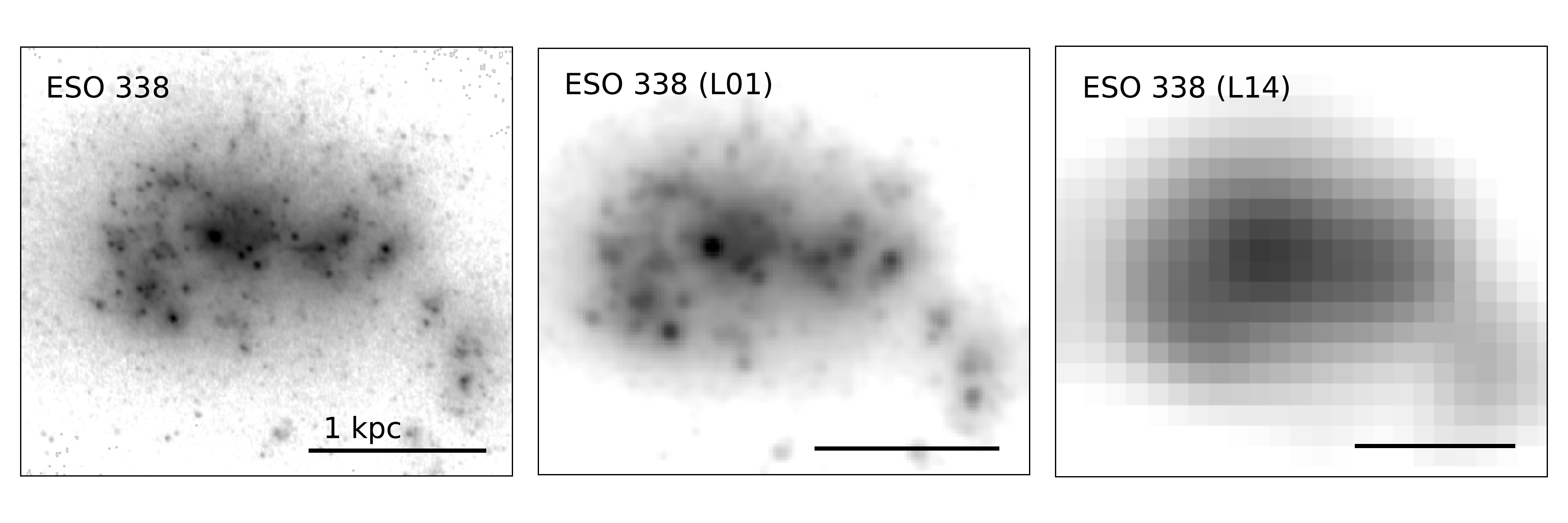}
\caption{ESO 338 observed in the F140LP filter (left panel) and its simulation at the redshift of LARS01 (central panel) and of LARS14 (right panel). North is up and east to the left. Black bars in the lower right corner show a physical size of 1 kpc.}
\label{fig:eso338_f140lp}
\end{figure*}

In order to simulate ESO 338 at different redshifts we convolve the images with the Gaussian kernel that produces a PSF with a FWHM of the correct physical scale (60 and 400 pc). After the convolution, the data frames are also re-binned in order to have a physical pixel scale corresponding to the ones of LARS01 and LARS14. Flux conservation is ensured throughout the steps. The original ESO 338 observation in the F140LP filter and the simulated frames at LARS01 and LARS14 redshifts are shown in Fig.~\ref{fig:eso338_f140lp}. For each of the frames we have repeated the clump extraction and analysis, with the following results:
\begin{itemize}
\item We extracted 157 sources (cluster candidates) in ESO~338, 2 of which have $\rm r_c>3.5$ px and are therefore excluded from the sample. Of the remainder, 139 were detected with magnitudes brighter than $30$ mag in more than 3 filters and constitute the final catalogue of the galaxy. We consider this the reference sample for our study as it has a sufficient resolution to contain genuine single clusters.  
\item Out of these, 57 sources were detected with a magnitude uncertainty lower than $0.3$ mag in at least 4 filters and are therefore considered part of the high-fidelity ($HF$) sub-sample. Differently from the LARS galaxies, the requirements for clumps to be part of the $HF$ sub-sample are applied to at least 4 filters (instead of requiring 5 filters) because in ESO 338 the F218W data are underexposed, leading to larger uncertainties and non-detections \citep[see][]{ostlin1998}. 
\item In the simulated ESO 338 at the distance of LARS01, which we will refer to as ESO 388 (L01) hereafter, we extracted 82 sources, 4 of which with $\rm r_c>3.5$ px. Using the same criteria as described above, 69 sources are part of the final catalogue and 26 are part of the $HF$ sub-sample.
\item In ESO~338 simulated at the LARS14 distance, i.e. ESO~388 (L14), we extract only 6 clumps, all with $\rm r_c<3.5$ px. 4 of them make the final catalogue and all 4 meet the requirements for being part of the $HF$ sample.
\end{itemize}
The physical resolution greatly affects the analysis of star-forming clumps in the galaxy. At larger distances, associations counting tens of clusters and clumps can only be studied as single sources. 

In the following sections we will characterize more in detail this bias, using the sizes and luminosities of the clumps extracted from the different realisations of ESO 388 datasets.
All the retrieved properties of ESO 338 clumps used in the following sections are summarized in Tab.~\ref{tab:eso338_prop}
\begin{table}
\centering
\caption{Results of the clump analysis of ESO 338 at different redshifts. The values reported in the rows are (in order): distance of the galaxy; range of effective radii detectable and median \reff of the sample; median luminosity (in F140LP filter); median clump SFR surface density; clumpiness parametrisations ($\rm F_{tot},\ F_{HF}$ and $\rm F_{3B}$ are described in the text in Section~\ref{sec:clumpiness}). $UV$ luminosities are in units of ($\rm 10^{41}\ erg/s$) while SFR surface densities, $\rm \Sigma_{SFR}$, in units of ($\rm M_\odot/yr/kcp^2$). Values within parentheses refer to the $HF$ sub-sample.}
\label{tab:eso338_prop}
\begin{tabular}{lrrr}
\hline
\ & ESO 338 & ESO 338 (L01) & ESO 338 (L14) \\
\hline
$\rm D\ (Mpc)$ & 37.5 & 119.1 & 645.9 \\
$\rm R_{eff, range}\ (pc)$  & $3.8 - 44.1$ & $9.4 - 110.2$ & $68.0 - 793.5$ \\
$\rm R_{eff, med}\ (pc)$  & $8.9\ (9.0)$ & $21.3\ (31.2)$ & $115.7\ (115.7)$ \\
$\rm L_{UV, med}$ & $0.2\ (0.7)$ & $0.6\ (2.8)$ & $31.1\ (31.1)$ \\
$\rm \Sigma_{SFR,med}$ & $3.4\ (7.1)$ & $1.3\ (2.3)$ & $3.5\ (3.5)$ \\
$\rm F_{tot}$   & $0.429$   & $0.568$   & $0.266$ \\
$\rm F_{HF}$    & $0.371$   & $0.460$   & $0.266$ \\
$\rm F_{3B}$    & $0.017$   & $0.044$   & $0.048$ \\
\hline
\end{tabular}
\end{table}

\section{Results}
\label{sec:results}
\subsection{Sizes}
\label{sec:sizes}
\begin{figure}
\centering
\subfigure{\includegraphics[width=\columnwidth]{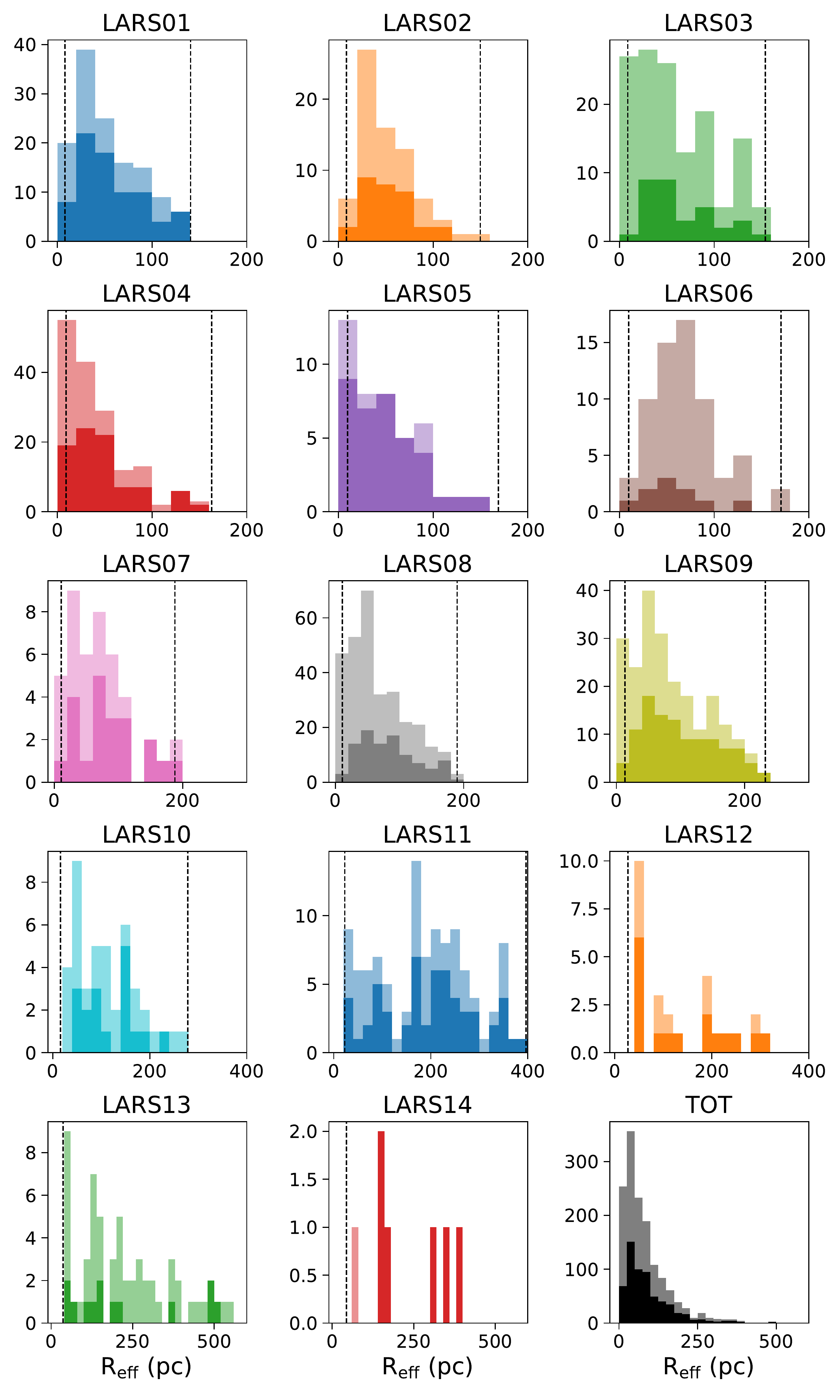}}
\subfigure{\includegraphics[width=\columnwidth]{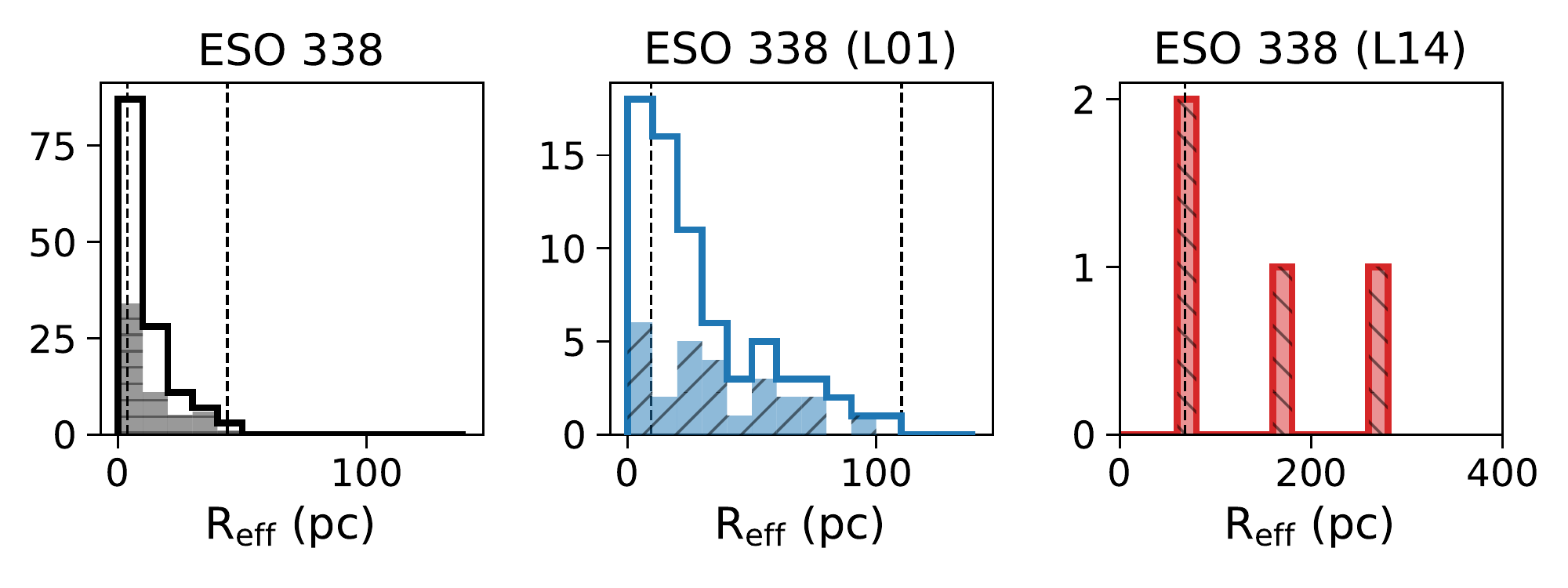}}
    \caption{Distribution of effective radii for all the LARS galaxies. Shaded bars represent the total sample, while full colours are used for the $HF$ sub-samples. Dotted vertical lines shows the minimum and maximum \reff\ detectable in each galaxy. The last line show the distribution of $\rm R_{eff}$ for the ESO 388 reference sample and its simulations at the redshifts of LARS01 and LARS14: in this case the contoured histogram refers to the total sample and the hatched shaded histogram refers to the $HF$ sub-sample.}
    \label{fig:sizes}
\end{figure}

The recovered size distributions of clumps for all LARS galaxies and for the ESO~338 reference sample are plotted in Fig.~\ref{fig:sizes}.
Due to our limitations in recovering radii smaller than 0.3 pixels described in the previous section, we have a number of unresolved sources in our sample, which in the figure have been assigned the \reff\ corresponding to $\rm r_c=0.3$ px.
There are 166 unresolved sources in the total sample and 37 in the $HF$ sample. In the closest galaxies the limiting \reff\ is of the order of 10 pc. The majority of the sources, however, are resolved and span a range that extends up to $\sim600$ pc. The median \reff\ value for the total sample is 58 pc, with 1st and 3rd quartiles at 33 pc and 108 pc (median of 71 pc for the $HF$, with quartiles at 39 pc and 122 pc). The ranges of \reff\ (in pc-scale) corresponding to the \rc\ range set by the limits at 0.3 px and 3.5 px are listed in Tab.~\ref{tab:size} along with the median values of the distributions in each galaxy.
The decrease in the number of clumps at larger sizes, observed in most of the galaxies, is, in part, an effect of the lower completeness for those systems, due to their lower surface brightness for a given luminosity (the completeness of the sample is described in Appendix~\ref{sec:test_completeness}).

The main trend retrieved in the size distributions across the LARS galaxies is the increase in the median clump size with redshift, caused by the decrease in physical resolution. This effect is clear in the analysis of clumps in ESO 338, whose sizes are plotted in the last row of Fig.~\ref{fig:sizes}. 
As summarized in Tab.~\ref{tab:eso338_prop}, in the original reference sample of ESO 338 we probe sizes in the range $\rm R_{eff}=3.8-44.1$ pc and we are therefore able to resolve most of the (single) star clusters. This ability is partially lost in ESO 338 (L01), where only the larger clusters (minimum $\rm R_{eff}=9.8$ pc) are probed. In ESO 338 (L14) we only detect a few systems, with minimum detectable effective radius $\rm R_{eff}=68.0$ pc, which is larger than the maximum $\rm R_{eff}$ at the original redshift. 

In nearby galaxies star cluster sizes have been studied in detail. Their typical size ranges between $1-10$ pc \citep[e.g.][]{lada2003,ryon2015,ryon2017} even if the most massive ones ($\rm \gtrsim10^6\ M_\odot$), usually called super star clusters, can extend up to $\sim20$ pc \citep[e.g.][]{meurer1995,bastian2013}. 
We deduce that the unresolved sources that we are observing in the closest LARS galaxies, with $\rm R_{eff}\sim10$ pc, may be dominated by single star clusters, while the majority of the sources detected, in the size range $\rm r_c=20-600$ pc are associations of clusters, usually called star cluster complexes \citep[e.g.][]{bastian2005}. 
The test done on the reference sample of ESO~338 reveals that many of these larger ($\rm r_c\gtrsim20$ ) clumpy star-forming regions contain multiple single clusters.
Therefore, the difference in the scales studied across the LARS galaxies implies a physical difference. The compact ($\rm r_c\sim10$ pc) sources observed in the closest LARS galaxies are possibly gravitationally bound clusters, able to survive and evolve within the host galaxy. On the other hand, many larger sources are possibly gravitationally unbound structures which will dissolve rapidly. Lacking dynamical information of the clumps we are unable to quantify this difference.
\begin{table}
\centering
\caption{Number of clumps, range of \reff\ values observable and median clump \reff\ for the LARS galaxies. Values between parentheses refer to the $HF$ sub-sample. The \reff\ range was calculated by converting the 0.2 and 3.5 px \rc\ limits (see text) into \reff\ values in pc units.} 
\label{tab:size}
\begin{tabular}{llll} %
\hline
Gal.	& $\rm N_{clumps}$	& \reff\ range	& \reff\ med. 	\\
\		& \ 				& (pc)				 & (pc) 	\\
\hline
L01	& 130 (78)	& $12.1-140.7$	& 42 (48)	\\ 
L02	& 73 (30)	& $12.8-149.8$	& 46 (50)	\\ 
L03	& 138 (33)	& $13.2-154.1$	& 49 (57)	\\ 
L04	& 163 (87)	& $14.0-163.0$	& 33 (40)	\\ 
L05	& 43 (36)	& $14.5-169.0$	& 41 (48)	\\ 
L06	& 65 (10)	& $14.6-170.5$	& 66 (54)	\\ 
L07	& 43 (21)	& $16.1-188.0$	& 65 (76)	\\ 
L08	& 305 (98)	& $16.3-189.7$	& 53 (78)	\\ 
L09	& 222 (107)	& $19.9-232.0$	& 69 (90)	\\ 
L10	& 43 (17)	& $23.9-278.5$	& 106 (111)	\\ 
L11	& 108 (56)	& $34.0-396.4$	& 187 (205)	\\ 
L12	& 26 (16)	& $40.3-470.3$	& 102 (116)	\\ 
L13	& 59 (13)	& $55.0-642.2$	& 206 (158)	\\ 
L14	& 7 (6)		& $65.3-761.6$	& 164 (232)	\\ 
TOT	& 1425 (608)	& $12.1-761.6$	& 58 (70)	\\ 
\hline
\end{tabular}
\end{table}

\subsection{Luminosity functions}
\label{sec:luminosities}
We study the clump magnitude distribution via the luminosity function. The luminosity function, defined as the number of sources per luminosity interval, $dn/dL$, has been extensively explored in the studies of young star clusters and HII regions in nearby galaxies. It is parametrized by a power-law, $dn/dL\propto L^\alpha$, whose slope $\alpha$ was observed to vary from galaxy to galaxy, spanning the entire range from $\alpha=-1.8$ to $\alpha=-2.8$ \citep[see][and references therein]{larsen2006}. Some studies suggest that the slope of the function could be correlated with properties of the host galaxy \citep{whitmore2014} or with different environments within single galaxies \citep{messa2018a,messa2018b}.
In some galaxies the power-law slope was observed get steeper at brighter magnitudes \citep{whitmore1999,gieles2010}, which is why the luminosity function is sometimes described by a Schechter function, with a exponential cut-off at bright magnitudes \citep{haas2008}. With few assumptions, the shape of the luminosity function can be used as a proxy for the mass distribution and this makes it a powerful tool in studying clusters. Even if the age-dependent light-to-mass ratio of clusters causes the luminosity function to not necessarily have the same shape as the mass function, if the cluster is a continuous power law with the same index at all ages the luminosity function will be a power law with the exact same index \citep{gieles2009}.

At larger physical scales, \citet{cook2016} studied tens of thousands $FUV$ star-forming regions in 258 nearby ($\rm D\leq 11$ Mpc) galaxies observed with \textit{GALEX}. Those regions span physical scales between $\sim20$ and $\sim500$ pc, similar to our sample. For every galaxy a clump luminosity function was derived and the resulting power-law slopes ($\alpha$) span a wide range of values, $-2.8 < \alpha < -1.0$, with a median value $\alpha=-1.83$. The slopes were studied in function of several galaxy properties but weak correlations were found only with SFR and SFR density ($\Sigma_{\textrm{SFR}}$).

A possible evolution of the clumps luminosity function with redshift was explored in \citet{livermore2012,livermore2015}, studying HII regions (observed in the \ha\ line) of physical scales $\sim100-1000$ pc in the redshift range $z=0-5$. The authors assumed for the luminosity function a fixed slope of $-1.75$, taken from the mass function slope of giant molecular clouds in \citet{hopkins2012} simulations, and derived a characteristic truncation luminosity which evolves with redshift, going from L$_0$(\ha)$\sim10^{41}$ erg/s at $\rm z=0$ to L$_0$(\ha)$\sim10^{43}$ erg/s for $\rm z>3$ clumps.

\begin{figure*}
\centering
\subfigure[LARS 01-12]{\includegraphics[width=\columnwidth]{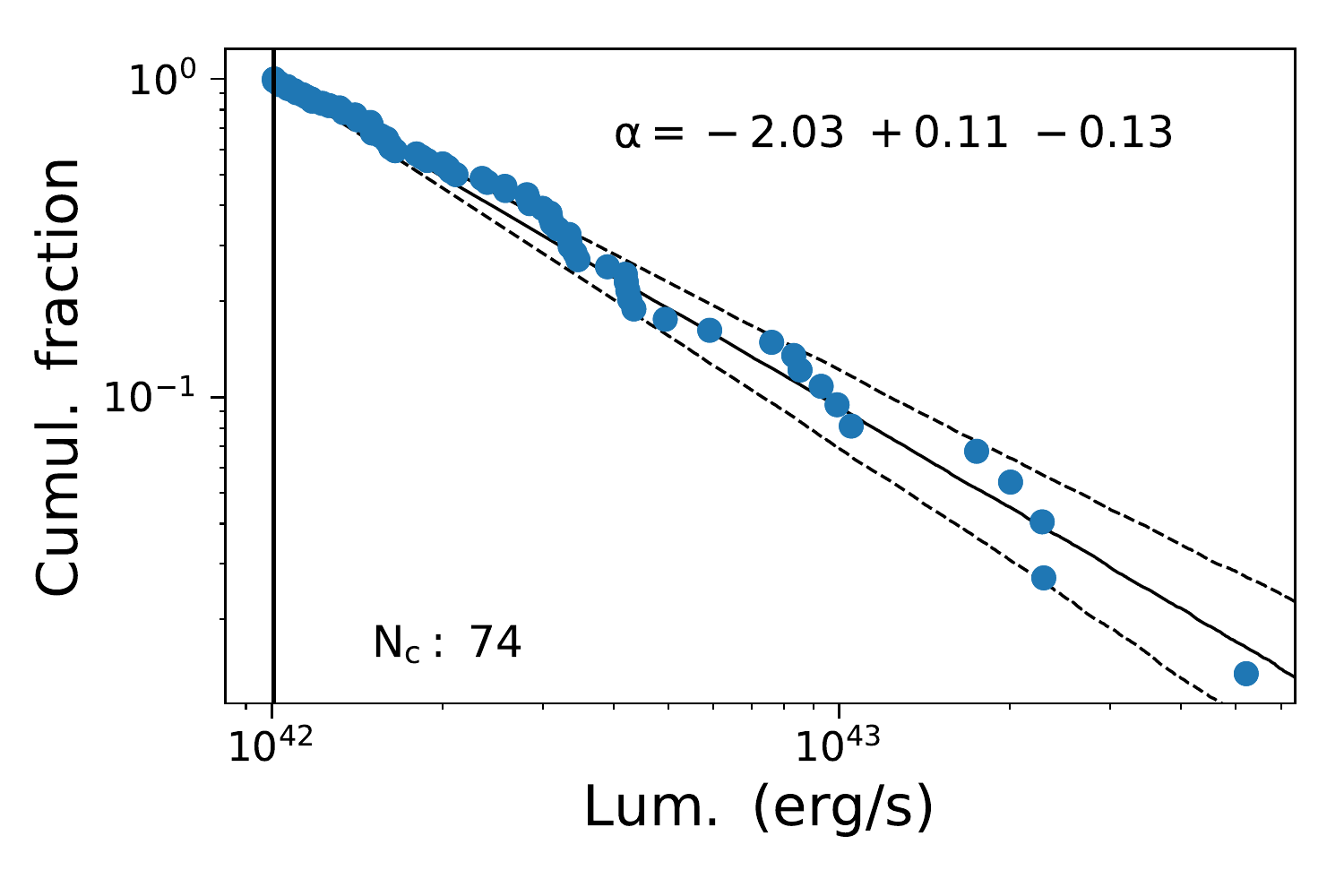}}
\subfigure[LARS 01-09]{\includegraphics[width=\columnwidth]{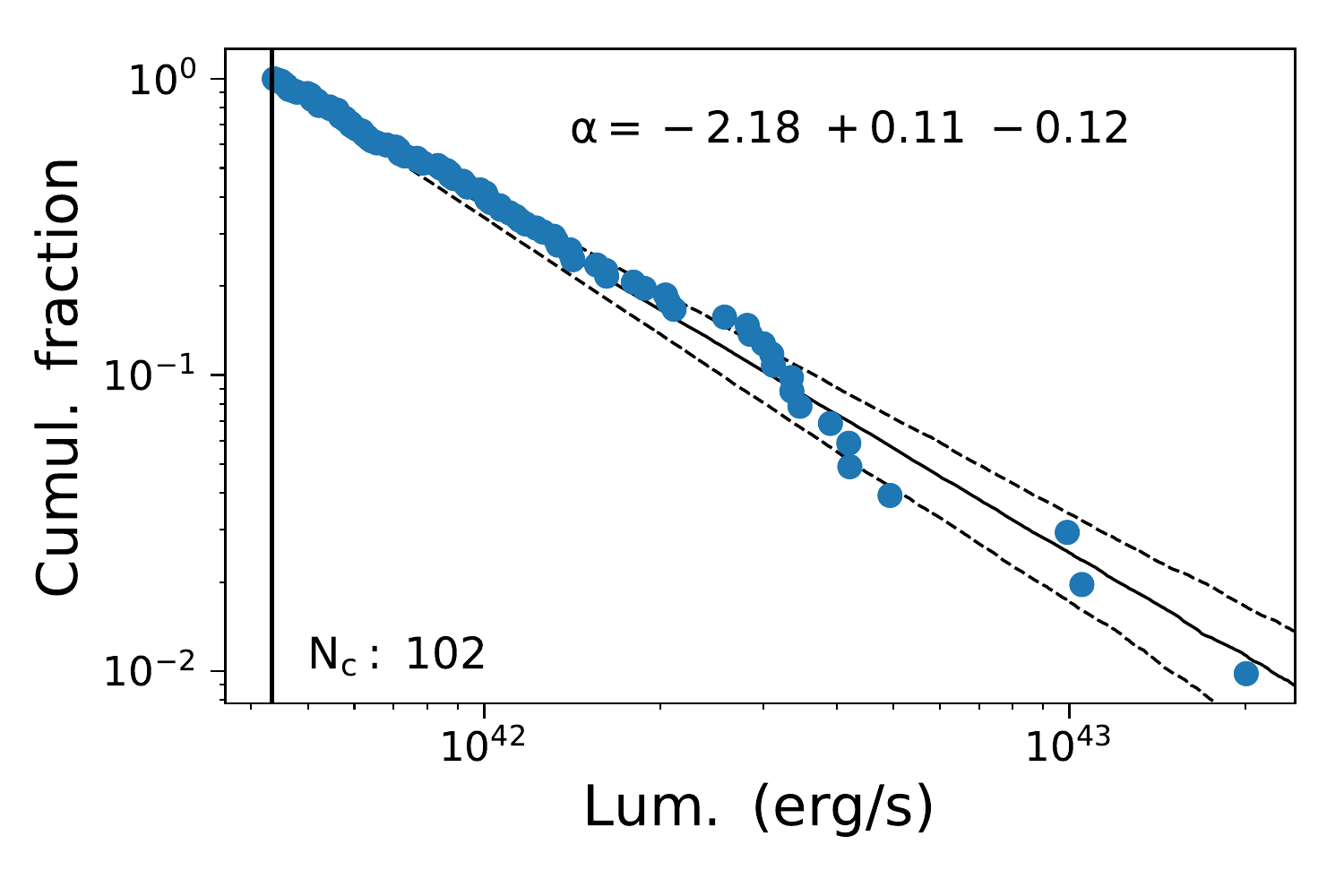}}
\caption{The cumulative luminosity function of the clumps in the galaxies from LARS 1 to 12 (panel a) and from LARS 1 to 9 (panel b). In both cases the lower luminosity limit (black vertical line) was derived using the most conservative value of all the galaxies considered (see text). The number of clumps in the function and the best-fit slope for a power-law are also reported. The solid and dashed lines are the best fit slope and the $\pm1\sigma$ slopes, respectively.}
\label{fig:lumfunc}
\end{figure*}

In the study of the LARS sample we are limited by the small number of sources per galaxy. We therefore build a global luminosity function including all clumps together. With the goal of focusing on the young (i.e. bluer) clumps and of comparing them to studies of star-forming clumps in literature, we chose to study their luminosity in the 1500 \AA\ $UV$ rest-frame (in detail, using the F140LP filter in all galaxies except LARS 13 and 14, where we used the filter F150LP due to their higher redshift).
Critical for the analysis of the luminosity function is the choice of the lower luminosity limit considered, which in turn relies on a good understanding of the magnitude completeness limits. As described in Appendix~\ref{sec:test_completeness}, LARS galaxies have different completeness levels in the $UV$. We chose the value at which completeness goes below $90\%$ as the lower magnitude limit for each galaxy. This limit is then converted into a luminosity and the most conservative value among all galaxies is used as the lower luminosity value for the global function. 
The value chosen is $\rm L_{lim}=10.08\times10^{41}$ erg/s which is the completeness limit of LARS11.The lower limits of the LARS galaxies are listed in Tab.~\ref{tab:lumfunc_lim}. The global luminosity function is plotted in Fig.~\ref{fig:lumfunc}(a). We did not include the clumps of LARS13 and LARS 14 from this analysis because the number of clusters hosted in these galaxies is small and the completeness limit is high. In total, there are 74 clumps above the completeness limit chosen. 
\begin{table}
\centering
\caption{$90\%$ completeness limits in $UV$ magnitudes (from Tab.~\ref{tab:completeness}) converted into luminosity limits (for the LARS galaxies used in the analysis of the luminosity function). We report also the number of sources above the luminosity completeness limits set by LARS07 and LARS11 ($\rm lim_{L07}$ and $\rm lim_{L11}$ respectively).}
\label{tab:lumfunc_lim}
\begin{tabular}{lllllll}
\hline
\ & L01 & L02 & L03 & L04 & L05 & L06 \\ 
\hline
$\rm mag_{lim}\ (mag)$  			& 22.0 & 23.0 & 24.0 & 24.0 & 23.0 & 24.0  \\
$\rm Lum_{lim}\ (10^{41}\ erg/s)$ 	& 2.44 & 1.11 & 0.49 & 0.54 & 1.38 & 0.59  \\
$\rm N\geq lim_{L07}$			    & 17    & 6 & 5 & 5 & 15 & 3 \\
$\rm N\geq lim_{L11}$			    & 7 & 3 & 1 & 2 & 8 & 1 \\
\hline
\ & L07 & L08 & L09 & L10 & L11 & L12 \\ 
\hline
$\rm mag_{lim}\ (mag)$ 			    & 22.0 & 24.0 & 23.0 & 24.0 & 23.0 & 24.0  \\
$\rm Lum_{lim}\ (10^{41}\ erg/s)$ 	& \textbf{4.34} & 0.82 & 3.15 & 1.64 & \textbf{10.08} & 5.76  \\
$\rm N\geq lim_{L07}$			    & 12    & 15 & 24 & 7 & 48 & 14 \\
$\rm N\geq lim_{L11}$			    & 5 & 3 & 11 & 5 & 19 & 9 \\
\hline
\end{tabular}
\end{table}
We assume a power-law shape for the luminosity function, $dn/dL\propto L^{\alpha}$, and therefore define the normalized probability of finding a source of luminosity $L_i$ as
\begin{equation}
p(L_i|\alpha,L_{\textrm{lim}}) \equiv \frac{L_i^\alpha\ \Theta(L_{\textrm{lim}})}{\int_{L_{\textrm{lim}}}^{\infty}L^\alpha\ \textrm{d}L}
\end{equation}
Using Bayes' theorem we know that the posterior distribution function for the slope $\alpha$ is 
\begin{equation}
p(\alpha|\{L_i\},L_{\textrm{lim}}) \propto \prod_{i}p(L_i|\alpha,L_{\textrm{lim}}),
\end{equation}
where $\{L_i\}$ is the observed luminosity distribution of Fig.~\ref{fig:lumfunc}(a).
We sample the posterior distribution of $\alpha$ using the \texttt{emcee} package (see Section.~\ref{sec:photometry}). The median value of the distribution is $\alpha=-2.03\ ^{+0.11}_{-0.13}$. The uncertainties stated are retrieved from the $\rm 84^{th}$ and $\rm 16^{th}$ percentiles of the distribution.
The slope is within the range of values found in the sample of \citet{cook2016} and in the studies of star clusters and HII regions in the local universe. 
The global luminosity function studied includes clumps that span a wide range in sizes, from $\rm R_{eff}<10$ to $\rm R_{eff}\sim$600 pc, and therefore a wide range of different objects, from clusters to extended star-forming regions.
In order to restrict the study of the luminosity function to smaller clumps we re-perform the analysis keeping only the galaxies from LARS01 to LARS09. In this way the largest clumps included have $\rm R_{eff}\sim200$ pc. We also notice that these galaxies are the ones hosting the most numerous clump populations. The luminosity function is shown in Fig.~\ref{fig:lumfunc}(b). Note that the lower luminosity limit was re-adapted to the selection, becoming $\rm L_{lim}=4.34\times10^{41}$ erg/s. There are 102 clumps above this completeness limit.
The best-fit value for the slope in this case is $\alpha=-2.18\ ^{+0.11}_{-0.12}$, which is consistent within uncertainties with the previous result.

In order to account for the low number statistics and to understand how the scatter of points may be affecting the results of the fit, we take into consideration two additional methods for estimating the uncertainties:
\begin{description}
    \item[``Jackknife'' method:] we remove one of the clumps from the sample and re-fit the luminosity function obtaining a new value for the slope. This is repeated for all the sources and we consider the median and the standard deviation of the resulting distribution of slopes as indicative of the best value and uncertainties.
    \item[Monte Carlo sampling of uncertainties:] We re-sample the luminosity of each source from a normal distribution centered on its value and with a standard deviation given by the magnitude uncertainty and we fit the new luminosity function. We repeat this process 1000 times. We consider the median and the standard deviation of the resulting distribution of slopes as indicative of the best value and uncertainties.
\end{description}
In both the cases just mentioned it would be computationally expensive to run the sampling of the posterior distribution, as done previously. We decide therefore to fit the function with a least-squares method. The fitted function in this case is the cumulative one, i.e.:
\begin{equation}
    y_{\textrm{cumul.}}\equiv \frac{N(>L)}{N_{tot}} \propto L^{\alpha+1}
\end{equation}
The results of these two additional methods are reported in Tab.\ref{tab:lumfunc_fit}. In both cases the uncertainties recovered are smaller than the ones found with the sampling of the posterior distribution.

\begin{figure}
\centering
\includegraphics[width=\columnwidth]{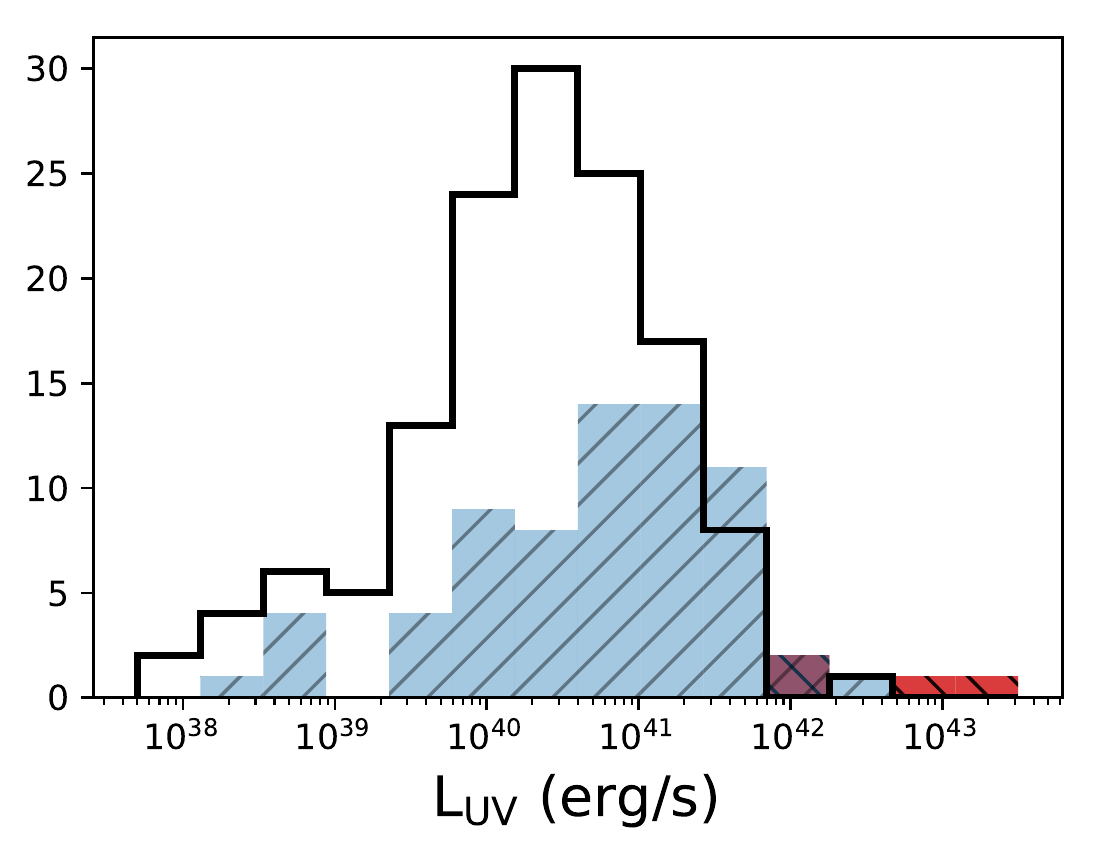}
\caption{Luminosity distribution of the clumps in ESO 338 (black contour), ESO338 (L01) (blue hatched) and ESO (L14) (red hatched). The plot shows that degrading the resolution causes the distribution to move to higher luminosities.}
\label{fig:e338_lum}
\end{figure}

The effect of resolution on luminosities was tested with the ESO 338 sample. As expected, at decreasing resolutions single clusters are merged together and the distribution of derived luminosities is shifted to brighter values (Fig.~\ref{fig:e338_lum}). 
We derive observed-magnitude completeness limits in $UV$ for ESO 338 and ESO 338 (L01) in a similar way to what done for the LARS galaxies (see Appendix~\ref{sec:test_completeness}), retrieving $\rm m_{lim}=20.5$ in both cases. This magnitude limit corresponds to limits in luminosity of $\rm L_{lim}=1.3\cdot10^{41}$ erg/s for ESO 338 and of $\rm L_{lim}=1.4\cdot10^{42}$ erg/s for ESO 338 (L01).
We study the luminosity function above the luminosity limit finding a slope $\alpha=-2.27^{+0.27}_{-0.32}$ for ESO 338. We point out that there are only 18 clumps above the limit. 
In the case of ESO 338 (L01) only two clumps have luminosities above the completeness limit, implying that most of the sample is affected by incompleteness. We fit the luminosity function of ESO 338 (L01) down to $\rm L_{lim}=1.3\cdot10^{41}$ erg/s, where we observe the peak of the luminosity distribution in Fig.~\ref{fig:e338_lum}, finding a slope of $\alpha=-1.92^{+0.16}_{-0.19}$. The flattening of the slope, compared to the result at the reference redshift, is therefore caused by incompleteness. A similar flattening in function of decreasing resolution was found in the analysis of the clump mass function in high-redshift galaxies in \citet{dz2018}.

\begin{figure}
\centering
\includegraphics[width=\columnwidth]{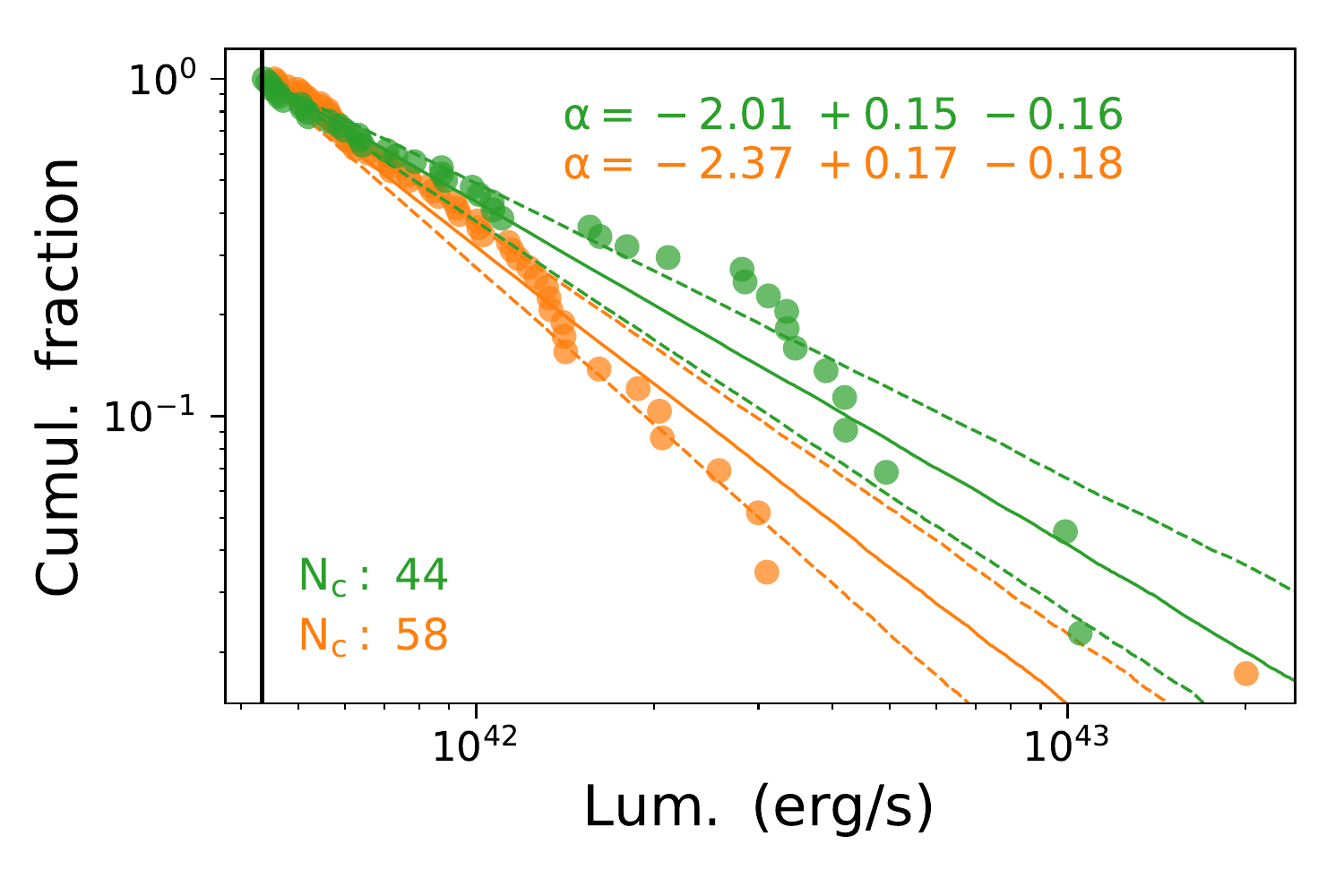}
\caption{Luminosity functions of LARS01 to LARS09 divided in two sub-samples with $\rm \Sigma_{SFR}>0.22\ M_{\odot}/yr/kpc^2$ (green) and $\rm \Sigma_{SFR}<0.22\ M_{\odot}/yr/kpc^2$ (orange). The slopes of the two functions differ by $\sim2\sigma$.}
\label{fig:lumfunc_Sigmasfr}
\end{figure}

As done by \citet{cook2016}, we investigate possible correlations between the luminosity function power-law slope and the SFR surface density of the host  galaxy. We again focus on the low-redshift galaxies of the sample (LARS 01 to 09) dividing them into two groups. We use the value of $\rm 0.22\ M_{\odot}/yr/kpc^2$ as the boundary for separating the two sub-groups because it results in groups of almost equal numbers of clumps (above the completeness value of $\rm L_{lim}=4.34\times10^{41}$ erg/s), 44 for the high-$\rm \Sigma_{SFR}$ sample including LARS01, LARS05 and LARS07, 58 for the low-$\rm \Sigma_{SFR}$ sample including the rest of the galaxies. We point out that the two groups contain galaxies at various redshifts.
The two luminosity functions are plotted in Fig.\ref{fig:lumfunc_Sigmasfr}, together with the best-fit slope values, $\alpha_1=-2.01^{+0.15}_{-0.16}$ and $\alpha_2=-2.37^{+0.17}_{-0.18}$. We recover slopes that differ by $\sim2\sigma$, and in particular a shallower slope for the high-$\rm \Sigma_{SFR}$ sample, consistent with what was found by \citet{cook2016}. This result can be interpreted as follows: galaxies with higher \sigmasfr\ (and therefore higher gas surface density, assuming the \citet{kennicutt1998} relation between $\rm \Sigma_{SFR}$ and $\rm \Sigma_{gas}$) form on average more luminous (and therefore more massive) clumps. 
An important caveat should be considered: all the LARS galaxies are on average highly star-forming galaxies, with large values of $\rm \Sigma_{SFR}$. The range of SFR densities that we are probing is therefore limited, and the comparison to galaxies with lower SFR densities (e.g. the galaxies of the eLARS sample, Melinder et al. (in prep)) could increase the significance of this result.
We test the dependence of the slope of the clump luminosity function on other galactic-scale properties, namely on $v_s/\sigma_0$ and on the galaxy stellar mass, $\rm M_*$. In the first case, we divide the galaxies in two samples using $v_s/\sigma_0=1$, commonly used to separate rotation-dominated galaxies from dispersion-dominated ones. In the second case we use $\rm M_*=2\times10^{10}\ M_\odot$ to separate the galaxies in two groups, since this value allow to have a similar number of clumps in the groups. We notice that, in our sample, most of the galaxies with high $\rm \Sigma_{SFR}$ also have low $v_s/\sigma_0$ and small stellar masses. As a consequence, we find that the slope of the clump luminosity function is shallower for low-mass and dispersion dominated galaxies (see Tab.~\ref{tab:lumfunc_fit}).

Assuming that a shallower slope indicates the presence of more massive clumps on average, we can try to put these result in the context of clump formation. As described by \citet{dekel2009}, the standard Toomre theory predicts that the typical clump mass scales with the disk mass of the host galaxy. In the case of LARS galaxies we find the opposite relation: the reason for this discrepancy can be attributed to the fact that the systems we are studying are highly perturbed, with a kinematics that is far from regular disks. We observe that in this case the formation of clumps may be regulated instead by the $\rm \Sigma_{SFR}$ and $v_s/\sigma_0$ parameters.
\begin{table}
\centering
\caption{Results of the fit of the clump luminosity function. The values in the columns are: (1) name of the sub-sample considered; (2) ID of the LARS galaxies included in the sub-sample; (3)-(6) luminosity function slope fitted with different methods described in the text: (3) via a MCMC sampling of the probability distribution, (4) least-squares fit of the cumulative function using a Jackknife method to retrieve the uncertainties, (5) least-squares fit of the cumulative function re-sampling the data via Monte Carlo methods.}
\label{tab:lumfunc_fit}
\begin{tabular}{llrrr}
\hline
\multicolumn{1}{c}{Sample} & \multicolumn{1}{c}{Included} & \multicolumn{1}{c}{$\alpha$} & \multicolumn{1}{c}{$\alpha_{\textrm{Jackknife}}$} & \multicolumn{1}{c}{$\alpha_{\textrm{MonteCarlo}}$}\\
\multicolumn{1}{c}{(1)} & \multicolumn{1}{c}{(2)} & \multicolumn{1}{c}{(3)} & \multicolumn{1}{c}{(4)} & \multicolumn{1}{c}{(5)}\\
\hline
$\rm Tot_{01-12}$           & 1-12          & $-2.03^{+0.11}_{-0.13}$   & $-1.94^{+0.02}_{-0.01}$   & $-1.95^{+0.04}_{-0.04}$ \\
$\rm Tot_{01-09}$           & 1-9           & $-2.18^{+0.11}_{-0.12}$   & $-2.11^{+0.02}_{-0.01}$   & $-2.12^{+0.04}_{-0.04}$ \\
$\rm low\ \Sigma_{SFR}$     & 2,3,4,6,8,9   & $-2.37^{+0.17}_{-0.18}$   & $-2.20^{+0.02}_{-0.02}$   & $-2.30^{+0.06}_{-0.05}$   \\
$\rm high\ \Sigma_{SFR}$    & 1,5,7         & $-2.01^{+0.15}_{-0.16}$   & $-1.91^{+0.04}_{-0.02}$   & $-1.87^{+0.04}_{-0.04}$   \\
low $v_s/\sigma_0$          & 2,5,7         & $-1.94^{+0.15}_{-0.18}$   & $-1.80^{+0.04}_{-0.03}$   & $-1.77^{+0.04}_{-0.03}$ \\
high $v_s/\sigma_0$         & 1,3,4,6,8,9   & $-2.36^{+0.15}_{-0.17}$   & $-2.24^{+0.02}_{-0.02}$   & $-2.32^{+0.06}_{-0.06}$ \\
low $\rm M_*$               & 1,2,5,6,7     & $-2.05^{+0.13}_{-0.15}$   & $-1.94^{+0.03}_{-0.02}$   & $-1.91^{+0.04}_{-0.03}$ \\
high $\rm M_*$              & 3,4,8,9       & $-2.40^{+0.19}_{-0.21}$   & $-2.25^{+0.03}_{-0.02}$   & $-2.37^{+0.07}_{-0.06}$ \\
\hline
\end{tabular}
\end{table}

As a last remark, we find that the luminosity function in the LARS galaxies is described by a simple power-law without the need of a truncation at high luminosities. The reason is likely the low number of clumps available: in order to sample the truncation a large statistical sample above completeness is necessary, usually of several hundreds of sources \citep[e.g.][]{Adamo2017}, which we are lacking. 

\subsection{Clumps SFR vs size relation}
\label{sec:surface_brightness}
We convert clumps' $FUV$ luminosities to SFR values using the relation in \citet{kennicutt2012}, assuming no intrinsic extinction. In a forthcoming paper, where we analyse the spectral energy distribution of clumps (Messa et al., in prep), we show that the majority of clumps have derived values of extinction $E_{B-V}<0.1$ mag. 
In Appendix~\ref{sec:a2} we show the effect of considering the extinction values derived in (Messa et al., in prep) on the clumps' SFR values.

\begin{figure*}
\centering
\subfigure[]{\includegraphics[width=0.49\textwidth]{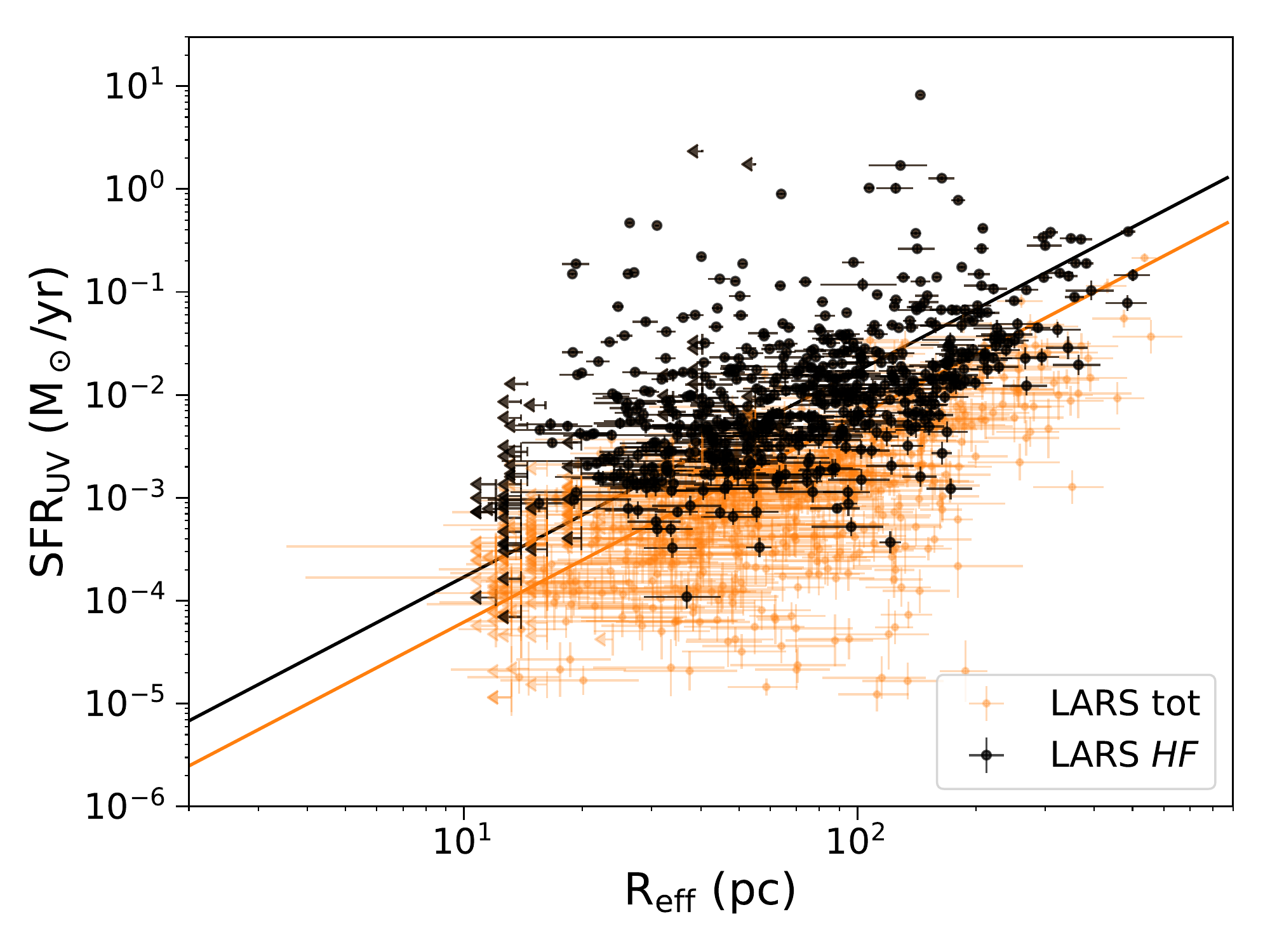}}
\subfigure[]{\includegraphics[width=0.49\textwidth]{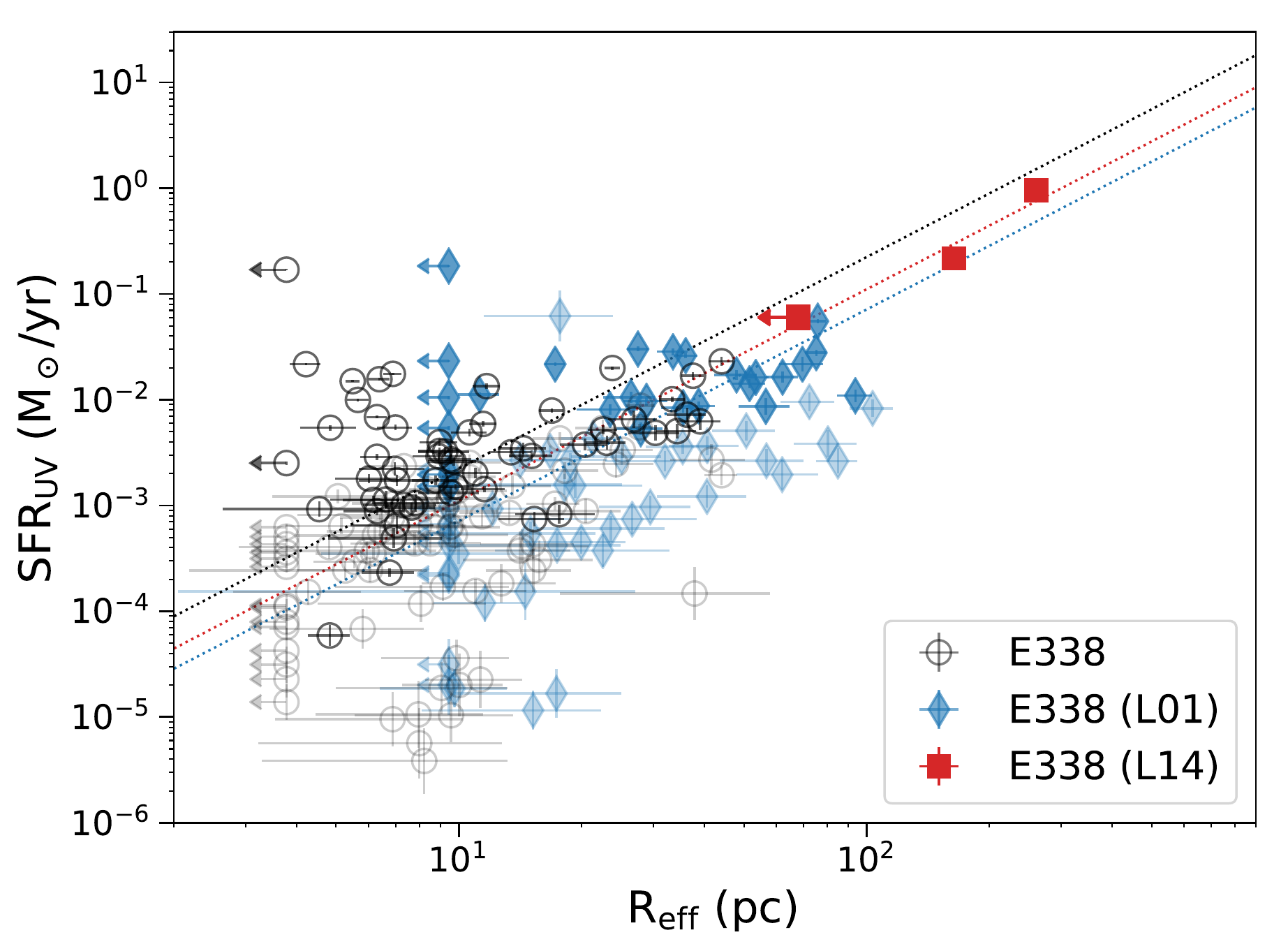}}
\subfigure[]{\includegraphics[width=0.49\textwidth]{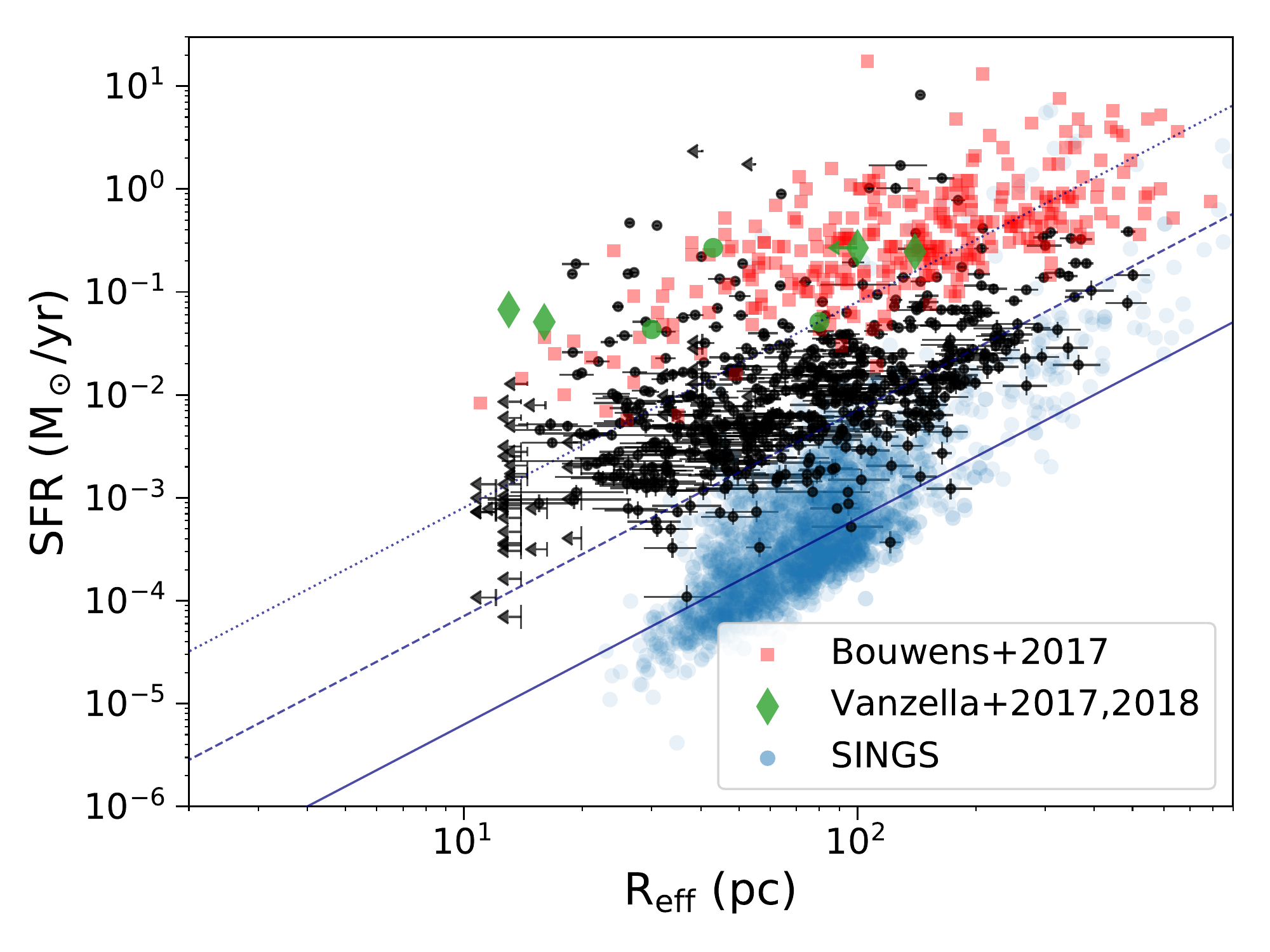}}
\subfigure[]{\includegraphics[width=0.49\textwidth]{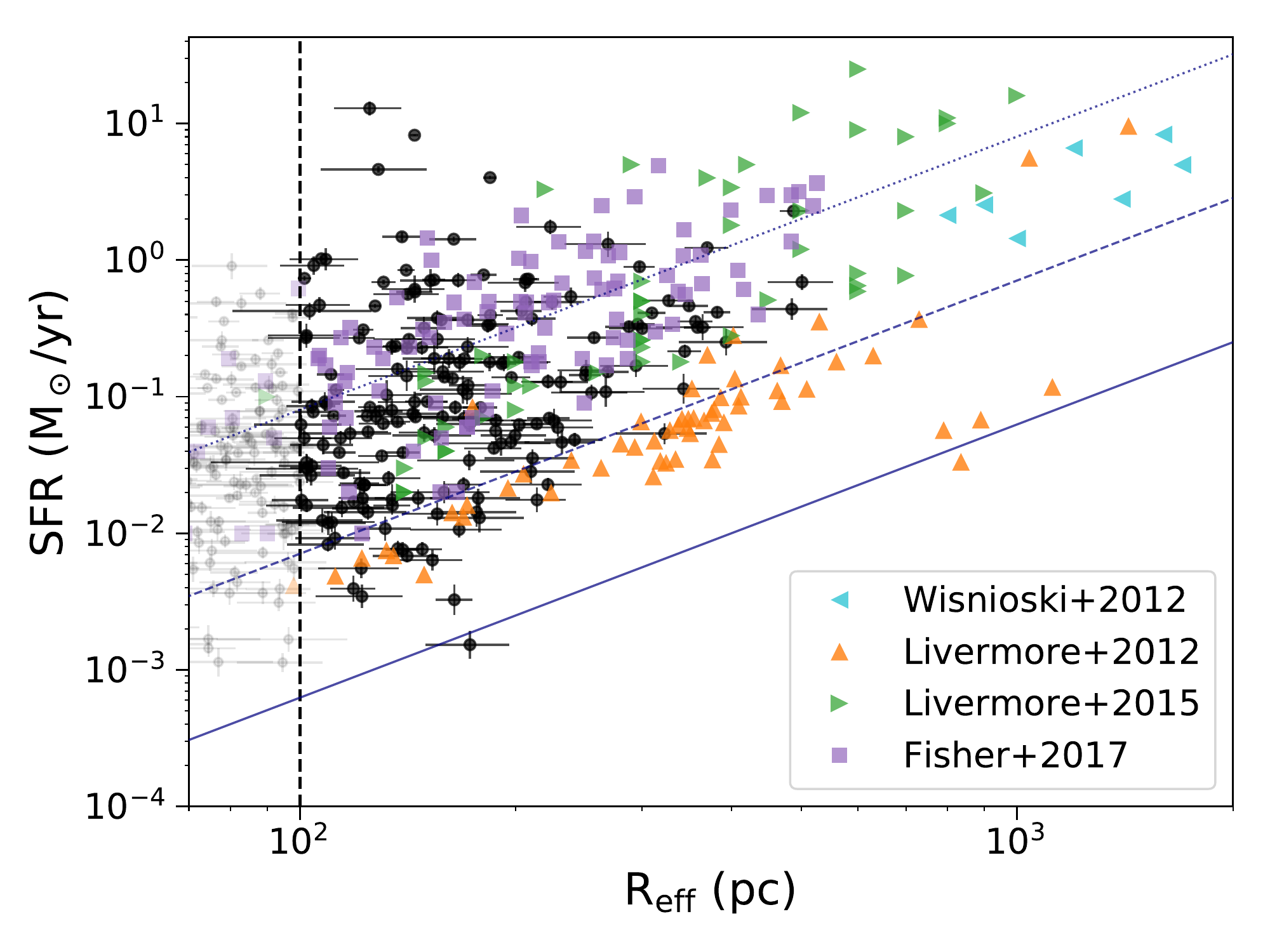}}
\caption{Size-SFR plots of the clumps in the LARS sample (a), ESO 338 at different resolutions (b) and comparison to other samples of clumps in literature (c and d). In panel (a) the total sample is represented by orange markers and the $HF$ sub-sample is in Black. Median values of \sigmasfr\ for the total and $HF$ samples are shown as orange and black solid lines respectively. In panel (b) the clumps in the reference sample of ESO 338 and in the convolved frames at L01 and L14 redshift are shown as black-contoured circles, blue diamonds and red squares, respectively. The $HF$ samples in this cases are plotted with solid colours, while the total samples have shaded colours. The dotted lines represent the median \sigmasfr\ values in the three cases. In panel (c) the $HF$ sub-sample of LARS clumps is compared to $\rm z=0$ clumps in SINGS galaxies \citep{kennicutt2003} as analysed in \citet{wisnioski2012} (blue circles) and to the samples of high-redshift galaxies from \citet{bouwens2017} (z$=6-8$, red squares) and \citet{vanzella2017b,vanzella2017,vanzella2018} ($z\sim3$ as green circles and z$\sim6$ as green diamonds). In panel (d) the $HF$ sample of LARS clumps with $\rm R_{eff}$ above 100 pc is compared to samples of \ha\ clumps at $\rm z\sim1.3$ from \citet{wisnioski2012} (cyan triangles), at $\rm z=1-1.5$ from \citet{livermore2012} (orange triangles), at $\rm z =1.5-4$ from \citet{livermore2015} (including data from \citealp{jones2010}, green triangles) and local galaxies at $\rm z\sim0.1$ from \citet{fisher2017} (purple squares). In panels c and d, the predicted \sigmasfr\ values from Eq.~\ref{eq:livermore15} at redshifts 0,1 and 3 are plotted as solid, dashed and dotted lines, respectively. In panels c and d the uncertainties on the values of the comparison catalogues are not reported for clarity of the plot.}
\label{fig:size_lum}
\end{figure*}

We show the SFR-size plot for our sample in Fig~\ref{fig:size_lum}(a). We consider both the total sample and the $HF$ sub-sample, and we notice that the clumps in the latter, at any radius, have on average higher values of SFR. As a consequence, the median $\rm \Sigma_{SFR}=SFR$/($\pi \textrm{R}_{\textrm{eff}}^2)$ of the $HF$ sub-sample is higher than for the total sample, as shown in Fig.~\ref{fig:size_lum}(a). This suggests, as expected, that the selection of the clumps with low photometric uncertainties implies a bias towards the clumps with higher SFR densities.

We analyse the effect of different redshifts using the clumps in ESO 338, in Fig.~\ref{fig:size_lum}(b).
At increasing simulated redshifts, the sources move towards the top-right corner of the plot, i.e. towards larger sizes and higher SFRs. In doing so, they may change their \sigmasfr: the sources in ESO 338 (L01) have a lower median SFR density than their counterparts at reference redshift. Only two sources on ESO 338 (L14) are not upper limits in $\rm R_{eff}$ and they have SFR densities compatible with the median of sources at the original redshift. 
We notice however that, by degrading the resolution, and therefore studying larger structures, we lose the possibility of characterizing the highest densities.
We conclude that the bright single clusters, when studied at larger scales will tend to blend with surrounding sources and have lower observed SFR densities: this result should be kept in mind when comparing clump studies at different redshifts and resolutions. 
This is consistent with the results of \citet{fisher2017}, who found that degrading the resolution and sensitivity of local clumps to match the resolutions reached in $\rm z=2$ lensed galaxies has the effect of moving the observed sample to lower \sigmasfr\ values. Part of this difference is caused by the apparent blending of many clumps into a single one, which causes the resulting clump to be brighter and larger in size, thus resulting in smaller \sigmasfr\ \citep[see][]{fisher2017}. A similar conclusion was also reached in \citet{tamburello2017} by degrading the resolution of \ha\ clumps from hydrodynamical simulations of clumpy disk galaxies. 

We compare LARS clumps to other samples in literature, taking care to compare clumps of similar physical scales. The SINGS sample \citep{kennicutt2003} contains local star forming galaxies (at distances $d<30$ Mpc), with HII regions resolved down to $\sim30$ pc size. In Fig.~\ref{fig:size_lum}c we plot their sizes and luminosities as derived in \citet{wisnioski2012}. 
In this comparison, it should be noted that SFRs for clumps in SINGS were derived using \ha\ observations (via the \citealt{kennicutt2012} relation), which are associated with slightly different time-scales (\ha\ probes on average younger emission than $UV$ bands) and sizes (\ha\ emission is associated to HII regions, while $UV$ radiation comes directly from the stellar objects). As for our sample, the SFRs of clumps in the SINGS sample were derived without accounting for intrinsic extinction.
We notice that the SINGS sample has on average lower SFRs than LARS $HF$ sample, despite the similar distribution in sizes. This difference points towards a real physical difference in density between the clumps in LARS and SINGS.

Recently, observations have been reported of lensed high-redshift galaxies where extremely compact star-forming regions have been detected, at scales down to $\sim10$ pc \citep{bouwens2017,vanzella2017b,vanzella2017,vanzella2018}. We take the absolute $UV$ magnitudes of those samples and convert them into SFR values as done for LARS.
Those high-redshift clumps have on average higher values of \sigmasfr than clumps in the LARS sample (see Fig.\ref{fig:size_lum}c); however there is some overlap between the samples, indicating that what was observed at high-redshift can be, as proposed, single star-forming regions or proto-globular clusters, which in some cases are so bright as to outshine the host galaxy.
The general difference between the SFR values of clumps in LARS and in those sample is not surprising: as mentioned in Section~\ref{sec:samples_comparison}, the selection of high-redshift galaxies in \citet{bouwens2017,vanzella2017b,vanzella2017,vanzella2018} is biased towards systems with extreme surface brightnesses (small radii and high intrinsic luminosities) that may not be representative of the clump population at z>3.

\citet{livermore2015} studied the redshift evolution of \sigmasfr\ using samples of \ha\ clumps at redshifts from $\rm z=0$ to $z\sim 5$ with sizes $\rm R\gtrsim100$ pc, suggesting that the mean surface brightness of star forming clumps evolve with redshift as:
\begin{equation}
\label{eq:livermore15}
\rm \log\left(\frac{\Sigma_{clump}}{M_\odot/yr/kpc^2}\right) = (3.5\pm0.5)\log(1+z)-(1.7\pm0.2)
\end{equation}
We plot in Fig~\ref{fig:size_lum}(d) the sizes and SFRs of the clumps studied in \citet{livermore2015}, together with their derived average \sigmasfr\ values at redshifts $\rm z=0$, 1 and 3, and compare them to the clumps with $\rm R_{eff}>100$ pc in the $HF$ sub-sample of LARS. In this case, since the SFRs of clumps in the comparison samples were derived taking into account external extinction, we also use for LARS extinction-corrected SFRs, as derived in Appendix \ref{sec:a2}.
Our sample covers a wide range in SFR densities, extending to much higher values than the predicted average for local galaxies. Most of the LARS clumps are found in the \sigmasfr\ range between $\rm z=1$ and $\rm z=3$ according to the \citet{livermore2015} prediction. Some clumps have even higher SFR surface densities, reaching the values found for $\rm z>3$ and partially overlapping with local ($\rm z\sim0.1$) clumps in the DYNAMO sample of high-redshift galaxies analogues \citep{fisher2017}. 

We try to understand the origin of this large scatter in clump SFR surface densities. 
Studying a compilation of \ha\ clumps from literature, \citet{cosens2018} showed that high and low-$\Sigma_{\textrm{SFR}}$ clumps follow different correlations in the $\rm R_{eff}-SFR$ space, possibly suggesting different origins (Str\"omgren spheres or star forming regions driven by Toomre instability). We divide our clumps sample in a high-SFR surface density ($\Sigma_{\textrm{SFR}}>1$ $\rm M_\odot/yr/kpc^2$) and a low-SFR surface density ($\Sigma_{\textrm{SFR}}\leq1$ $\rm M_\odot/yr/kpc^2$) sub-samples, similarly to what done in \citet{cosens2018}, and we fit a function of the form $\textrm{L}_{\textrm{UV}} \propto \textrm{R}_{\textrm{eff}}^\gamma$. We do not find a difference in the derived $\gamma$ slopes in the two sub-samples, with $\gamma=1.67\pm0.12$ for the high-$\Sigma_{\textrm{SFR}}$ sub-sample and $\gamma=1.66\pm0.06$ for the low-$\Sigma_{\textrm{SFR}}$ sub-sample (see Appendix~\ref{sec:a22} for more details on the fit). Both values are close to $\gamma = 1.74$ found for clumps with $\Sigma_{\textrm{SFR}}>1$ $\rm M_\odot/yr/kpc^2$ in \citet{cosens2018} and close to a relation $L\propto r^2$ expected for star forming regions driven by Toomre instability.
On the other hand, we observe that the SFR surface density of clumps do depend on the galactic-scale properties of their host galaxies. 
We divide the galaxies in sub-samples using the same division of Section~\ref{sec:luminosities} (Tab.~\ref{tab:lumfunc_fit}) and we calculate the median clump SFR surface density in each sub-sample. We find that clumps have on average higher SFR surface density in galaxies characterized by high $\Sigma_{\textrm{SFR}}$, low $v_s/\sigma_0$ and low $M_*$ (see Appendix~\ref{sec:a23} for more details). Similarly to what found in the study if the luminosity function, this result suggests that the galactic-scale properties of the host galaxies affect the clump $\Sigma_{\textrm{SFR}}$.

\subsection{Clumpiness}
\label{sec:clumpiness}
Recent studies of the redshift evolution of star-forming regions were motivated by the discovery that galaxies at high redshift appear more \textit{clumpy} than galaxies in the local universe in the rest-frame $UV$ \citep[e.g.][]{elmegreen2009}. 
In this section we study the \textit{clumpiness} of the LARS galaxies as a function of galactic-scale properties. We use two main parametrisations for the clumpiness:
\begin{enumerate}
\item Fraction of the galaxy $UV$ light in clump. This method simply express the clumpiness as the relative contribution of clumps to the galaxy UV emission \citep[e.g.][]{soto2017}.
\item Fraction of the galaxy $UV$ light in the brightest clump. In studying the galaxies of the CANDELS/GOODS-S and UDS fields in the redshift range $\rm z=0.5-3$, \citet{guo2015} showed that defining \textit{clumpy} galaxies as those where the brightest (off-centred) clump accounts for at least $8\%$ of the $UV$ light (rest-frame wavelength in the range $2000-2800$ \AA\ in their sample) allows to distinguish the high-z star-forming main-sequence galaxies from nearby spirals.
\end{enumerate}
We measure the total rest-frame $1500$ \AA\ $UV$ flux of the LARS galaxies inside the regions defined in Section~\ref{sec:lars_properties}.
In order to estimate clumpiness following method (i), we consider the clumps with \reff$<200$ pc and sum up their $UV$ flux. The ratio between the clumps $UV$ flux and the galactic $UV$ flux gives the first estimator of clumpiness, that we will call $\rm F_{tot}$ for the rest of the paper. We calculate this ratio considering only clumps in the $HF$ sub-sample ($\rm F_{HF}$). The size limit at \reff$=200$ was imposed to ensure that we are considering clumps at similar scales in all galaxies, as suggested by \citet{johnson2017}.
To parametrise clumpiness following method (ii), we considered the $UV$ flux of the 3rd brightest clump in each galaxy and divide it by the galactic $UV$ flux ($\rm F_{3B}$). We consider this measurements more solid than considering the brightest clump within each galaxy, as the latter may correspond with the nuclear region of the galaxy.
We present the clumpiness values of the LARS galaxies according to these parametrisations in Tab.~\ref{tab:clumpiness}.
\begin{table}
\centering
\caption{Clumpiness of the LARS galaxies according to different parametrisations, as described in the text.}
\label{tab:clumpiness}
\begin{tabular}{rrrr}
\hline
\multicolumn{1}{c}{\ } & \multicolumn{1}{c}{$\rm F_{tot}$}  & \multicolumn{1}{c}{$\rm F_{HF}$} & \multicolumn{1}{c}{$\rm F_{3B}$} \\
\hline
L01	& $0.68\ \pm0.01$	 & $ 0.66\ \pm0.01$	& $0.063\ \pm0.002$  \\
L02	& $0.66\ \pm0.01$	 & $ 0.62\ \pm0.01$	& $0.066\ \pm0.003$  \\ 
L03	& $0.40\ \pm0.01$	 & $ 0.37\ \pm0.01$	& $0.033\ \pm0.002$  \\
L04	& $0.31\ \pm0.01$	 & $ 0.28\ \pm0.01$	& $0.017\ \pm0.001$  \\
L05	& $0.65\ \pm0.01$	 & $ 0.65\ \pm0.01$	& $0.061\ \pm0.001$  \\
L06	& $0.57\ \pm0.01$	 & $ 0.41\ \pm0.01$	& $0.059\ \pm0.003$  \\
L07	& $0.81\ \pm0.01$	 & $ 0.79\ \pm0.01$	& $0.102\ \pm0.003$  \\
L08	& $0.44\ \pm0.01$	 & $ 0.37\ \pm0.01$	& $0.017\ \pm0.001$  \\
L09	& $0.44\ \pm0.01$	 & $ 0.40\ \pm0.01$	& $0.018\ \pm0.001$  \\
L10	& $0.47\ \pm0.01$	 & $ 0.41\ \pm0.01$	& $0.056\ \pm0.004$  \\
L11	& $0.13\ \pm0.01$	 & $ 0.11\ \pm0.01$	& $0.021\ \pm0.001$  \\
L12	& $0.73\ \pm0.02$	 & $ 0.73\ \pm0.02$	& $0.134\ \pm0.016$  \\
L13	& $0.56\ \pm0.01$	 & $ 0.52\ \pm0.01$	& $0.053\ \pm0.004$  \\
L14	& $1.00\ \pm0.01$	 & $ 1.00\ \pm0.01$	& $0.038\ \pm0.003$  \\
\hline
\end{tabular}
\end{table}

\begin{figure*}
\centering
\includegraphics[width=0.99\textwidth]{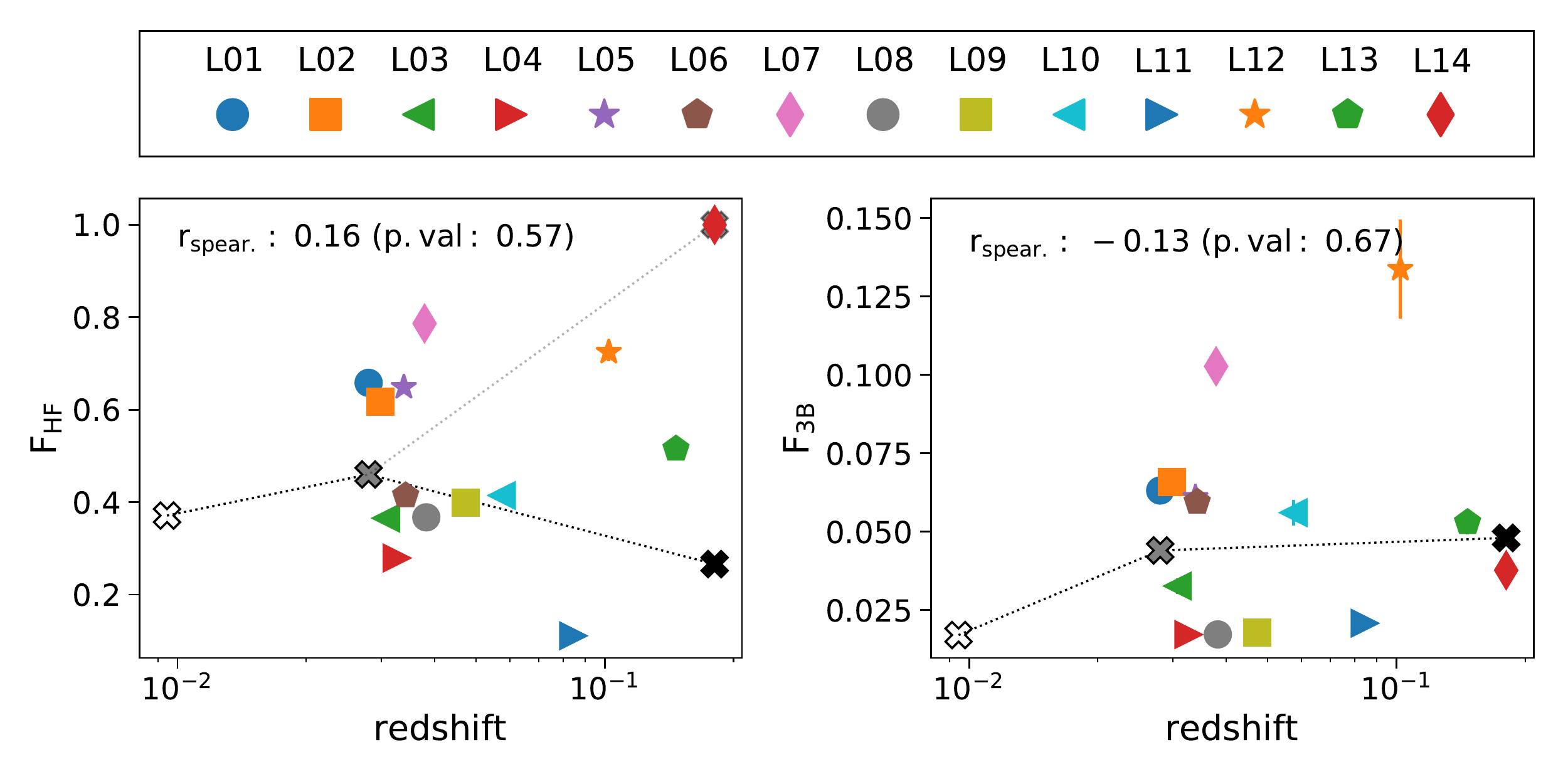}
\caption{Clumpiness parametrised as the fraction of the galaxy $UV$ light in clumps ($\rm F_{HF}$, left panel) and as the fraction of the galaxy $UV$ light in the $\rm 3^{rd}$ brightest clump ($\rm F_{3B}$, right panel) in function of redshift. The ``X'' symbols are the values of clumpiness for ESO 338 at its real redshift (white), at the redshift of LARS01 (grey) and at the redshift of LARS14 (black). In the left panel we also added the $\rm F_{HF}$ value that ESO 338 (L14) would have if clumps of all sizes were considered ($\rm F_{HF}=1$).}
\label{fig:clumpiness_redshift}
\end{figure*}
Clumpiness is shown as a function of redshift in Fig.~\ref{fig:clumpiness_redshift}. The clumps account for more than $50\%$ of the $UV$ flux in half of the LARS galaxies.
In Fig.~\ref{fig:clumpiness_redshift} we also added the clumpiness of ESO 338 at the three different redshifts considered in this study. 
The analysis of ESO\,338, shows that clumpiness changes by less than $20\%$ going from $z\approx0.01$ to $z=0.028$ and even declines when simulating the galaxy at $z=0.18$. Having imposed the restriction of only including clusters with sizes of $\rm R_{eff}<200$ pc, we avoid the effect of clumpiness increasing as a function of redshift: without this limitation, ESO\,338 simulated at $z=0.18$ would have a clumpiness value of $\rm F_{HF}=1$ as can be seen in the left panel of Fig.~\ref{fig:clumpiness_redshift}.
The same figure shows that the values of clumpiness in the LARS galaxies do not show a steady increase with redshift, suggesting that the resolution may be affecting the derived $\rm F_{HF}$ only in the most distant galaxies. The large scatter in the clumpiness values for LARS galaxies in their narrow redshift dynamic range confirms that the specific clumpiness of each galaxy is set by its internal properties more than by its resolution. We test the dependence of the clumpiness measurements on redshift by running a Spearman's rank correlation test. Correlation coefficients and their associated probabilities (p-values) are collected in the first column of Tab.~\ref{tab:spearman}. The coefficients are below 0.2, with high p-values, indicating no correlation.

We notice that LARS galaxies have a higher clumpiness that local galaxies. \citet{larsen2000} measured the fraction of U-band light contributed by young star clusters to the total U-band luminosity of the galaxy in a sample of local galaxies, finding that in spiral galaxies the median fraction is $0.5\%$. Even accounting for the increase of $\sim20\%$ observed in ESO338 when going from resolving clusters to clumps, the clumpiness of local spirals is still one order of magnitude below what we observe in the LARS galaxies.

We explore possible correlations of the clumpiness with the galaxy SFR and gas dynamics properties derived in Section~\ref{sec:lars_properties}. Specific SFR ($\rm sSFR$) and SFR surface density ($\rm \Sigma_{SFR}$) are derived dividing the $UV$-derived SFR by the stellar mass $\rm M_*$ and the galaxy area ($\pi \textrm{r}_{\textrm{g}}^2$) respectively. We run a Spearman's rank correlation test on each combination of clumpiness parametrisation-galaxy property. We note that the number of galaxies in our sample is limited, but we consider the results of the test as indicative of possible correlations. The results of the test are collected in Tab~\ref{tab:spearman}.
\begin{table*}
\centering
\caption{Results of the Spearman's rank correlation test for each combination of the clumpiness parametrisation and galactic-scale properties. For each combination both the correlation coefficient and the associated $\rm p-value$ (in brackets) are reported. The clumpiness parametrisations are described in the text.}
\label{tab:spearman}
\begin{tabular}{lrrrrrr}
\hline
\multicolumn{1}{c}{\ } & \multicolumn{1}{c}{redshift} & \multicolumn{1}{c}{$v_s/\sigma_0$} & 
\multicolumn{1}{c}{sSFR} & \multicolumn{1}{c}{\sigmasfr} & \multicolumn{1}{c}{$\rm M_*$} & \multicolumn{1}{c}{$\rm f_{esc}$(\lya)} \\
\hline
$\rm F_{tot}$	& $0.10\ (0.725)$	& $-0.71\ (0.005)$	& $0.45\ (0.110)$	& $0.65\ (0.012)$   & $-0.64\ (0.014)$   & $0.64\ (0.013)$ \\
$\rm F_{HF}$	& $0.16\ (0.573)$	& $-0.71\ (0.004)$	& $0.49\ (0.078)$	& $0.71\ (0.004)$   & $-0.56\ (0.039)$   & $0.67\ (0.009)$ \\
$\rm F_{3B}$	& $-0.13\ (0.670)$	& $-0.67\ (0.009)$	& $0.14\ (0.637)$	& $0.34\ (0.233)$   & $-0.58\ (0.030)$   & $0.53\ (0.051)$ \\
\hline
\end{tabular}
\end{table*}
\begin{figure*}
\centering
\includegraphics[width=1.\textwidth]{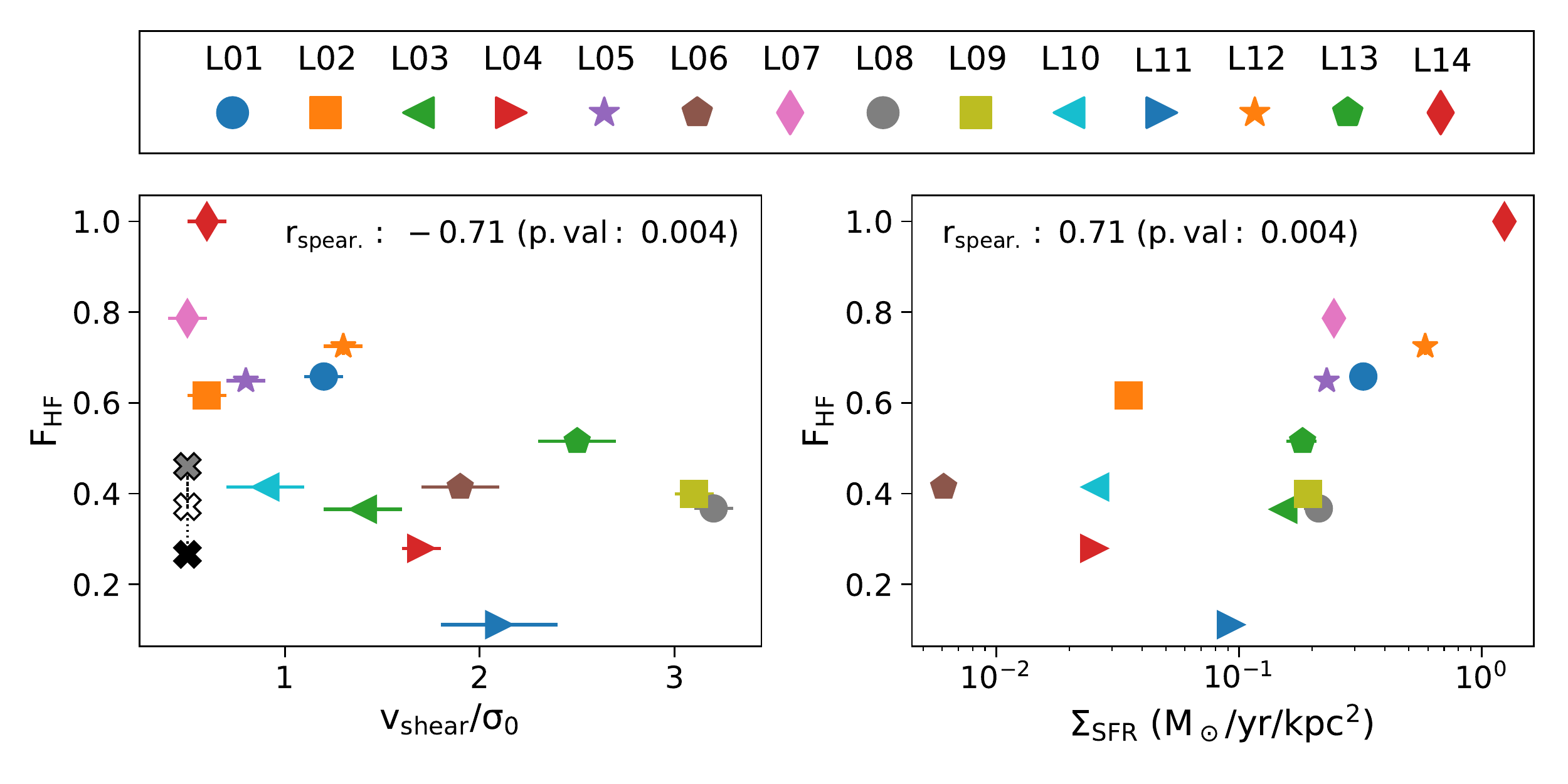}
\caption{Clumpiness measured for the clumps with $\rm R_{eff}<200$ pc in the $HF$ sub-samples in function of the shear over dispersion velocity (left) and of \sigmasfr\ (right). Spearman's rank correlation tests return  correlations of $-0.71$ (p: 0.004) for the left plot and of $+0.71$ (p: 0.004) for the right plot.}
\label{fig:clumpiness_correlations}
\end{figure*}
We observe an anti-correlation between clumpiness and \vssz\ and a positive correlation between clumpiness and the galaxy SFR surface density. The clumpiness shows only a weak correlation with specific SFR (sSFR) and a weak anti-correlation with the galaxy stellar mass, $\rm M_*$. We show the correlations between $\rm F_{HF}$ and both $v_s/\sigma_0$ and \sigmasfr\ in Fig.~\ref{fig:clumpiness_correlations}. 
The high level of scatter and the low number of galaxies in our sample poorly constrain the p-values associated with the correlation coefficients. We can ask ourselves how the correlation coefficient retrieved would change if LARS14 was removed from the sample. As stated before, at the redshift of LARS14 the clumpiness values derived could be affected by the poor physical resolution. Removing the points corresponding to LARS14 data in Fig.~\ref{fig:clumpiness_correlations}, the Spearman's test finds correlation coefficients of $\rm r_{spear}=-0.66\ (p.val:0.014)$ and of $\rm r_{spear}=0.64\ (p.val:0.019)$ for the left and the right plots, respectively. The correlation found proves that is not LARS14 alone to drive the derived correlations.

The correlations found in this section were similarly found for the DYNAMO sample in \citet{fisher2017}, where \ha\ clumps were proven to have a higher contribution to the host galaxy emission in galaxies which are more dispersion-dominated (lower $v_s/\sigma_0$) and have higher sSFR. 
The DYNAMO sample includes nearby ($z\sim0.1$) galaxies with high gas fraction and elevated gas velocity dispersions, resembling the properties of high-redshift galaxies \citep{fisher2017}.
While in DYNAMO the galaxies hosted a gas-rich rotation-supported disk, the same is not true for the LARS galaxies, which, both from morphological \citep{guaita2015} and dynamical \citep{herenz2016} studies appear to be mostly merging systems with irregular morphologies (only 2 out of 14 galaxies were classified as rotating disks in \citealt{herenz2016}). 
\citet{fisher2017b} used the clump properties of the DYNAMO sample to validate the disk instability models \citep[e.g.][]{dekel2009} that are expected to regulate star formation at high-redshift. The study of clumps in our sample suggests that, even if the star-formation event is driven by galaxy interactions, the clumpiness is affected by the SFR surface density of the galaxy (which we can consider as a proxy for the gas surface density) and by the rotational-over-dispersion velocity ratio in the gas. 

\subsubsection{Lyman-$\alpha$ escape fraction vs clumpiness}
In addition to the galactic-scale properties studied as a function of clumpiness in the previous section, we focus on the relation between clumpiness and the escape of \lya\ radiation. Clumps are the sources of most of the ionizing radiation (Messa et al. in prep.) and we also know that the escape of \lya\ radiation from galaxies is very dependent on the gas distribution at sub-galactic scales. We can therefore expect that galaxy morphology and clumpiness have an impact on the amount of \lya\ radiation escaping. It has been for example suggested that the \lya\ equivalent width of high redshift galaxies is related to their morphologies and sizes, with compact galaxies having larger equivalent widths compared to galaxies with more extended and diffuse emission or disks \citep{pentericci2010,cowie2011,law2012,Paulino-Afonso2018}. A similar result is found in low-redshift galaxies, where LAEs are found to have more compact morphologies compared to NUV-continuum selected galaxies \citep{cowie2010}.

\begin{figure}
\centering
\includegraphics[width=\columnwidth]{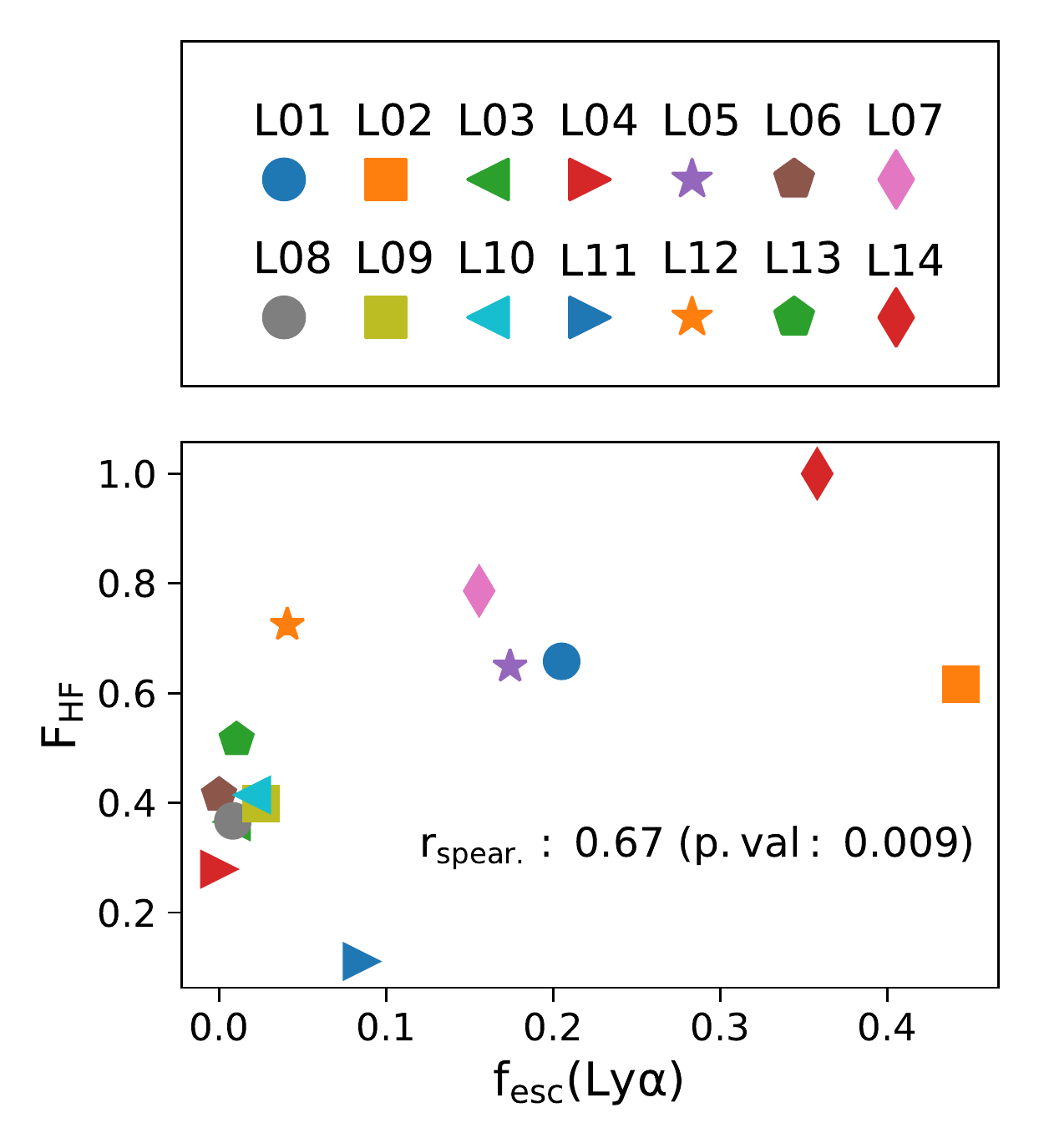}
\caption{Clumpiness of the $HF$ sub-sample in function of the \lya\ escape fraction for the LARS galaxies. Galaxies with $\rm f_{esc}$(\lya)$>0.1$ have a clumpiness higher than $50\%$. The coefficient and associated probability of the Spearman's rank test are 0.7 and 0.006 respectively.}
\label{fig:clumpiness_lya}
\end{figure}

\citet{hayes2014} presented a value of the \lya\ escape fraction, $\rm f_{esc}$(\lya), for each of the LARS galaxies. We use updated escape fraction from Melinder et al (in prep.) which are listed in Tab.~\ref{tab:lars}.
We ran a Spearman's rank correlation test between $\rm f_{esc}$(\lya) and the clumpiness parametrisations described in the previous section, finding, on average, a good correlation, the strongest being with $\rm F_{HF}$ (Tab.~\ref{tab:spearman}). We point out again that we have a limited number of galaxies in our sample and therefore the correlation we derive cannot have high statistical significance. When plotting the correlation (Fig.~\ref{fig:clumpiness_lya}), we notice that not all the galaxies with elevated clumpiness have a high fraction of \lya\ escape. It is instead true that all the galaxies with $\rm f_{esc}$(\lya)$>0.1$ have clumpiness higher than $50\%$. 

The escape fraction of \lya\ in the LARS galaxies was already studied in previous works and, for example, \citet{herenz2016} found a hint of anti-correlation between $\rm f_{esc}$(\lya) and \vssz. The anti-correlation, shown in the previous section, between \vssz\ and the clumpiness could therefore predict a clumpiness-$\rm f_{esc}$(\lya) correlation, consistent with what has been found here.

\section{Conclusions}
\label{sec:conclusions}
We report on the sizes and luminosity properties of star-forming clumps, in the LARS sample of nearby ($\rm z=0.03-0.2$) galaxies analogues of high-redshift Lyman-break galaxies.
The total sample counts 1425 sources, 608 of which have photometric uncertainties smaller than 0.3 mag in the 5 broad bands in which the galaxies are observed. Focusing on the $UV$ band (rest-frame wavelength $\sim1500$ \AA) we have investigated luminosity distribution, SFRs of the clumps, as well as their contribution to the total $UV$ emission of the host galaxy, which we refer to as \textit{clumpiness}. 
In doing so, we also consider the clumps properties as a function of galactic-scale properties, namely \sigmasfr and the ratio between the rotational and dispersion velocity of the galaxy gas ($v_s/\sigma_0$, measured from \ha\ observations in \citealp{herenz2016}).
We obtain the following results:
\begin{enumerate}
\item We find clump size between $10-600$ pc, similar to the range found in the literature for clumps in high-redshift galaxies.
\item The $UV$ luminosity function of the clumps in the LARS galaxies can be described by a power-law with slope $\alpha=-2.03^{+0.11}_{-0.13}$, similarly to what has been found in other nearby galaxies \citep[e.g.][]{cook2016}. When dividing the clump sample as a function of the host-galaxy \sigmasfr, we find that clumps in galaxies with higher SFR density have a shallower luminosity function, i.e. they are on average more luminous (and therefore more massive). The same is true in galaxies with low values of $v_s/\sigma_0$ and low stellar masses.
\item Converting the $UV$ luminosities of clumps into SFR values using \citet{kennicutt2012} relation, we study the size-SFR relation of our sample, and compare to other published samples, both local and at high redshift. We find that LARS clumps have on average a higher SFR density than clumps observed in $\rm z=0$ star-forming galaxies. Considering the redshift evolution of \sigmasfr\ of clumps suggested by \citet{livermore2015}, our sample is compatible with SFR density in galaxies at $\rm z=1-3$. Some clumps have extreme SFR surface densities, compatible with those found in galaxies at redshift beyond 3. The median SFR surface density of the LARS clumps is higher in galaxies with high \sigmasfr, low $v_s/\sigma_0$ and low $\rm M_*$.
\item LARS galaxies have $UV$ morphologies dominated by clumps. In many galaxies the clump contribution is $>50\%$ of the total $UV$ emission. We find indications of correlation between the clumpiness and SFR surface density and of anti-correlation with $v_s/\sigma_0$.
\item We find moderate positive correlation between the clumpiness and the \lya\ escape fraction: all the LARS galaxies with $\rm f_{esc}$(\lya)$>0.1$ have clumpiness higher than $50\%$. 
\item In order to account for the resolution effects caused by the different redshifts of galaxies in the LARS sample, we performed the same clump analyses in the nearby galaxy ESO 338-IG04, hosting a population of $>100$ star clusters at a distance of $37.5$ Mpc. The analyses were repeated degrading the resolution of ESO 338-IG04 in order to simulate its observation at the redshifts of LARS01 and LARS14 (the nearest and the most distant galaxies in the LARS sample). This test shows that degrading the resolution causes the clumps to appear larger and brighter, affecting also the study of clumps SFR surface densities: a better resolution allows the characterisation of star-forming regions with higher \sigmasfr. Finally, the galaxies appear more clumpy when imaged at lower resolution. This result should be kept in mind when comparing clumps studied at different scales, especially in high-redshift galaxies, where is usually difficult to constrain sizes lower than $\sim100$ pc.
\end{enumerate}
We conclude suggesting that the elevated star-formation conditions of the sample, probably set by mergers and interactions \citep{guaita2015,herenz2016}, can drive the formation of clumps with elevated surface brightnesses, that contribute to a large fraction of the $UV$ emission of the host galaxies. 
However, LARS covers a narrow dynamic range in SFRs. In the future, the inclusion of the eLARS sample (Melinder et al., in prep), consisting of galaxies in the same redshift range but with, on average, lower SFRs will help probing the effect of SFR on clumpiness.

\section*{Acknowledgements}
Based on observations made with the NASA/ESA \textit{Hubble Space Telescope}, obtained at the Space Telescope Science Institute, which is operated by the Association of Universities for Research in Astronomy, Inc., under NASA contract NAS 5-26555. These observations are associated with program \#12310,\#11522. The authors gratefully thank Dr. R.Bouwens and Dr. E.Wisnioski for sharing their catalogues. The authors are also thankful to the anonymous referee for comments and suggestions that helped improving the manuscript. A.A., G.\"O. and M.H. acknowledge the support of the Swedish Research Council (Vetenskapsr\r{a}det) and the Swedish National Space Agency (SNSA). M.H. acknowledges the continued support as Fellow of the Knut and Alice Wallenberg Foundation.




\bibliographystyle{mnras}
\bibliography{biblio} 




\appendix

\section{Tests of the photometric fitting analysis}
\label{sec:a1}
We report here additional details and testing of the photometric analysis described in Section~\ref{sec:photometry}.
The test included in this section are:
\begin{enumerate}
\item choice of the box-size for fitting (Section~\ref{sec:a11});
\item test of the size resolution (Section~\ref{sec:lowersize});
\item test of the effect of non-detections in some filter (Section~\ref{sec:a13});
\item test of the recovered size when considering 5 filters or only the F140LP filter (Section~\ref{sec:a15});
\item test of the impact of source subtraction after each source's fitting (Section~\ref{sec:a14});
\item test of the completeness limits of each galaxy (Section~\ref{sec:test_completeness}).
\end{enumerate}

\subsection{Choosing the background and box size}
\label{sec:a11}
\begin{figure}
\centering
\includegraphics[width=\columnwidth]{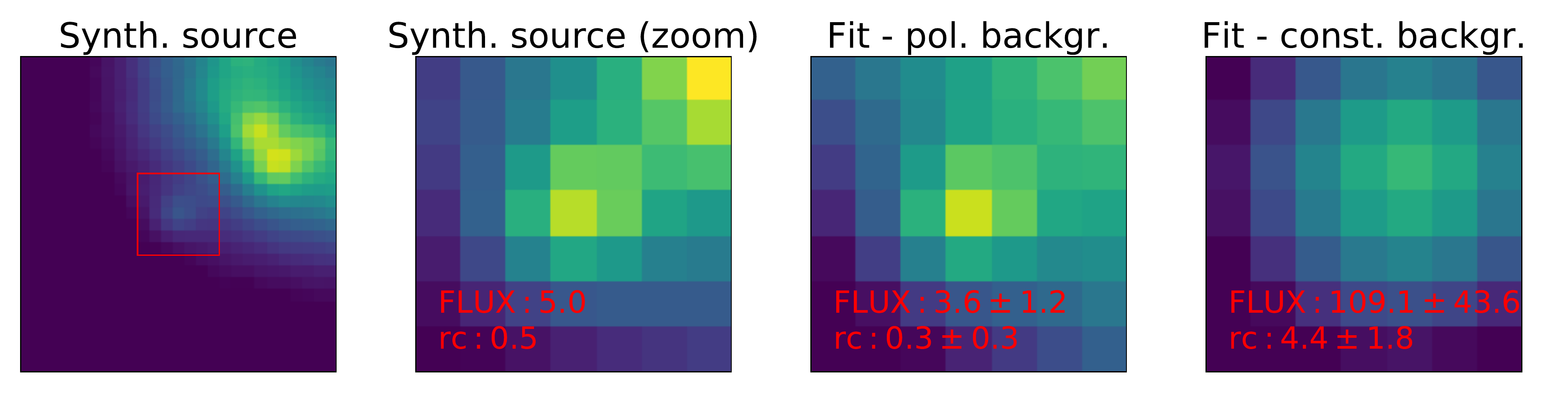}
\caption{Test fit of a synthetic clump inserted in an region of LARS01 characterized by strongly varying background (left and centre-left panels). The fit with a $1^{st}$ degree polynomial background (centre-right) is able to recover consistent values, while the fit with uniform background cannot produce reliable results.}
\label{fig:test_background}
\end{figure}

The two major problems encountered in performing the photometric analysis were the crowding of the sources and the contamination from background emission. They put strong constraints on the details of the fitting routine, in particular on the choice of the number of parameters used to model the observation and the fitting radius.
We modelled separately the clumps and the background. Clumps were modelled by Moffat profiles convolved with the instrumental PSF. The background was found to change on very small spatial scales. While testing the code we recognized that modelling it with a constant could lead to large errors, even when the area considered for the fit is small. As an example, we show in Fig.~\ref{fig:test_background} the fit of a synthetic source placed in a region of LARS01 with no close neighbouring sources but with a strongly varying background. A fit with a background modelled as a $\rm 1^{st}$ degree polynomial gives a good results, while when the background is modelled as a constant value we retrieve completely wrong values for size and flux. This simple example explains why it is important to accurately model the background. We also found that in such small regions (scales of $\sim7$ pixels), considering polynomials of higher degrees does not improve the quality of the fit (while instead increasing the number of free parameters of the fit). 

The choice of the number of parameters used for the fit (4 for the clump, 3 for the background) and the sizes of the sources to be fitted strongly influence the choice of the area considered in the fit. We computed the fit in a box region centred on the source, which should be large enough to include the core region of wide sources and whose number of pixels is supposed to be large enough to be able to constrain all free parameters. At the same time the box should be as small as possible in order not to include too much contamination from nearby sources. We performed tests adding synthetic sources of different radii and magnitudes in the $B-$band frame of LARS01, and using cut-out boxes of different sizes to fit them. In general, we found that increasing the box area increases the quality of the fit. We decided to consider a $7\times7$ px box as the standard dimension for our fitting process because is a good compromise between being able to retrieve the properties of large sources ($\rm r_c\approx 2-3.5$ px) and not introducing nearby sources in the fit area. For input core radii in the range $\rm r_c=0.3-3.5$ and input magnitudes between 20 and 25 mag, the fit in the $7\times7$ px box results in radii and magnitudes that differ in $50\%$ of the cases by less than 0.14 px and 0.15 mag respectively.

We point out that keeping the same pixel size for the fitting box in all the galaxies implies a different physical scale according to the redshift (7 pixels correspond to 163 pc at the distance of LARS01 but corresponds to $\sim650$ pc at the distance of LARS14).  
The choice was motivated by the ability of characterizing the $\rm r_c$ in the same way across different galaxies. The consequence of studying sources with different physical sizes was already explored in the text in Section~\ref{sec:sizes}.

\subsection{Testing the lowest detectable size}
\label{sec:lowersize}
\begin{figure}
\centering
\includegraphics[width=\columnwidth]{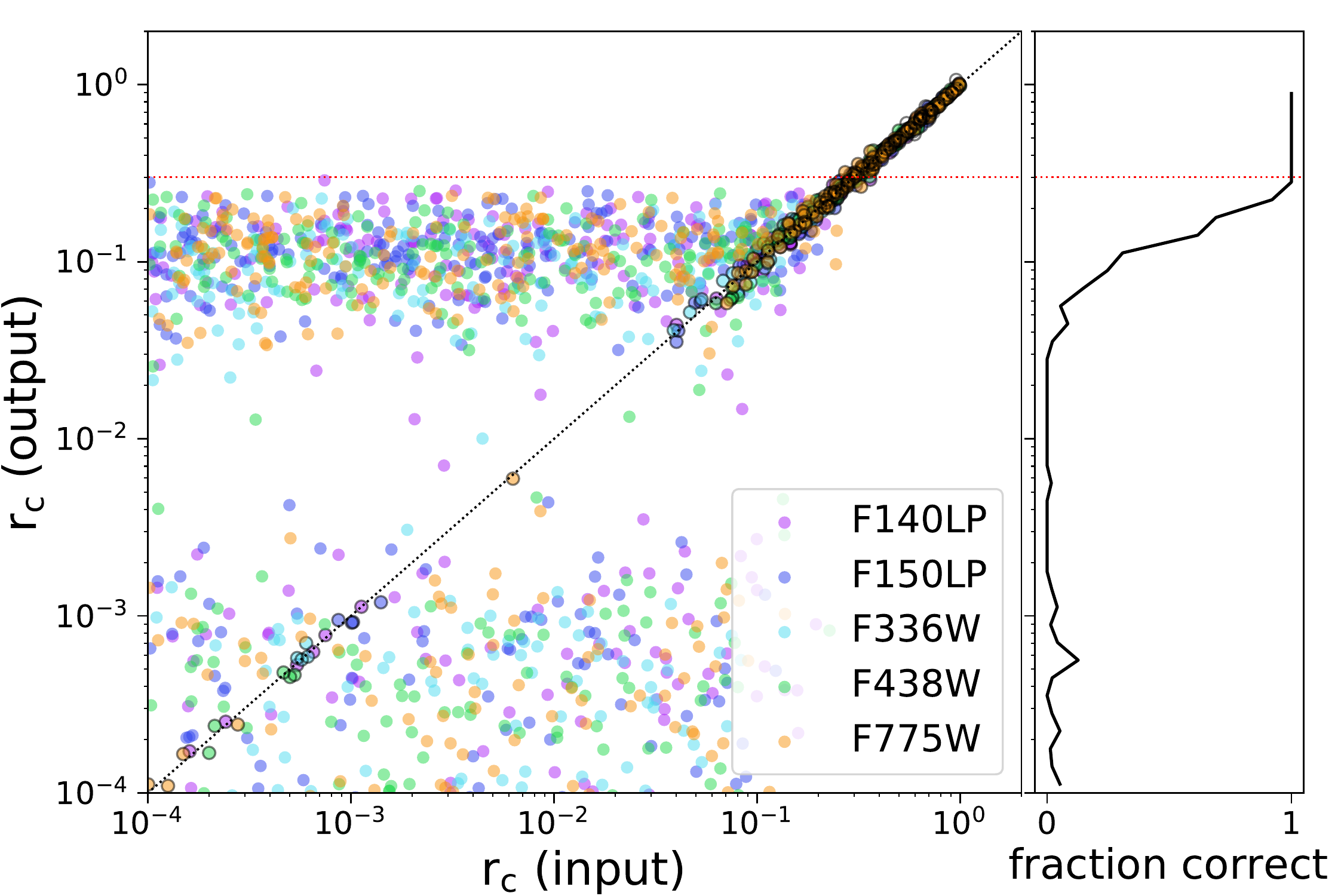}
\caption{Test for the fit of PSF-like sources. (Left panel): comparison between the simulated sizes, \rc\ (input), and the values derived from the fit, \rc\ (output). The 1 to 1 relation is represented by a black dotted line and sources whose difference between output and input \rc\  are less than $20\%$ of their input size are denoted by a black edges. (Right panel): fraction of the  correctly derived sizes (within $20\%$ of the input value) in function of the output size. We see that all sources for which we derive $\rm r_c\geq0.3$ have been correctly derived, while our fitting method is not sensitive to lower $\rm r_c$ values.}
\label{fig:test_rcmin}
\end{figure}

The fitting routine models the observed clump shapes as convolution between the instrumental PSF and a Moffat function of core radius \rc\ (which parametrizes the intrinsic radius of the source) and for this reason is in principle able to derive infinitely small \rc\ value. In practice, however, extremely small \rc\ values are not distinguishable. To test the lowest \rc\ value that we can confidently distinguish with our code, we simulate synthetic clumps with \rc\ values uniformly distributed in logarithmic space between $10^{-4}$ and $1$ px, putting them at random positions inside LARS01 galaxy and running the photometry code to derive their size. Each filter and was treated independently and only one source at the time was simulated, in order to avoid crowding effects. In total 2500 sources were simulated (500 per filter), with luminosities between 23 and 20 mag. Fig.~\ref{fig:test_rcmin} shows the results of the test: all sources with detected size above 0.3 px have a size difference with the input \rc\ of less than $20\%$. Below that value the code starts to be insensitive and the output \rc\ values are distributed randomly in comparison to the input values. Following the results of this test we assign a value $\rm r_c=0.3$ px to all clumps with retrieved \rc\ values below that value, and treat them as upper limits.

\subsection{Non-detection in some of the filters}
\label{sec:a13}
\begin{figure}
 \centering
	\includegraphics[width=0.85\columnwidth]{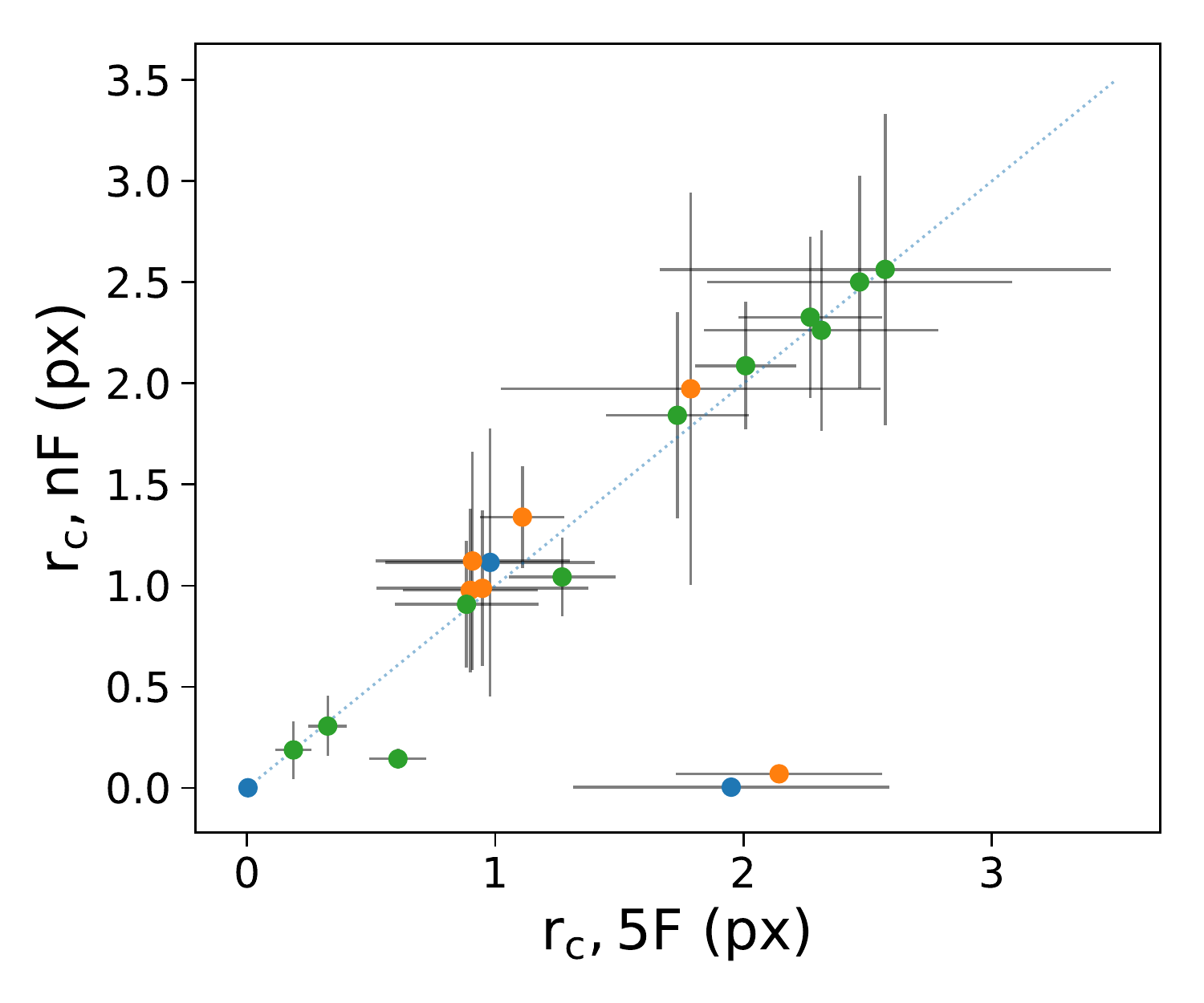}
    \caption{Comparison between the \rc\ values retrieved considering all the 5 filters (5F), or only the filters where the source is detected (nF) in LARS01 sources detected in 4 bands (green), 3 bands (orange) or 2 bands (blue).}
  \label{fig:test_nfilters}
\end{figure}

Clumps were extracted in the $B$ band, without requiring a detection in the other filers. For this reason in some cases the photometry routine could be affected by the fact that a source is not detected in some of the 5 bands considered. An incorrect size derivation of the source caused by this non-detection would in turn affect the flux measurement in filters where the source is detected. We tested the effect of non-detections in LARS01 by running the photometry routine a second time on clumps which have a non-detection ($\rm mag>30$) in at least one filter. This time photometry is run considering only the filters where the source was detected. A comparison between the fit in all 5 filters and in the detection-filters only is given in Fig.~\ref{fig:test_nfilters}.
In most of the cases the derived sizes are the same within the uncertainties. In three cases (out of 20) the two results are incompatible, meaning that the inclusion of all filters have disturbed the fit. For this reason during the analyses of the current paper we selected also a sub-sample of clumps detected in all 5 bands.
We point out that most of the source have a non-detection only in one filter out of five.

\subsection{Size fit only in the F140LP filter}
\label{sec:a15}
In our work we derived the size of the clumps using the 5 filters at the same time. This is fundamental because for each clump we want to study the emission from the same region in all filters, in order to be consistent when using the multi-band photometry to fit the broad-band spectral energy distribution (Messa et al., in prep.). 
Since in this current paper we mostly focused on the clumps emission in the UV band, we tested how the size derivation would be different if only filter F140LP was considered in the size-photometry fit. For this test we used the F140LP map of LARS01 before the convolution to a common PSF. We find that the median difference in the derived core radii is $0.1$ px, with a standard deviation of $0.9$ px. This result suggests that we are not introducing a bias in the size measures (the $\rm R_c$ derived using 5 filters is only $0.1$ px larger, on average). 
We consider the fit in 5 bands more stable against the background variations and crowding effects. We expect that both these effects are reduced when more than one filter is considered.

\subsection{Testing the source subtraction}
\label{sec:a14}
The photometry routine we developed fits sources starting from the brightest to the least bright, where this `preliminary' brightness was measured in the $B$-band via aperture photometry. After the fit of each source, our routine subtracts its best-fit model (source only, not background) from the science frame, in order to avoid the contamination from the tails of bright sources to the flux of nearby dimmer ones in subsequent fits. To measure the effect of this source subtraction over the derived sizes and magnitudes, we ran on LARS01 a photometric analysis which is identical to the one discussed in the text, except for the fact that clumps fluxes were not subtracted from the science frames after the fitting. Results are given in Tab.~\ref{tab:test_subtraction}. The median difference in the recovered \rc\ values is $-0.02$ px, with first and third quartiles of the distribution being $\pm0.21$ px. The difference in \rc\ values does not depend on the magnitudes, but it affects them. However, the difference in recovered magnitudes in the 5 filters have median values very close to zero. The scatter in the distributions depends on the filter but on average in $50\%$ of the cases are within $\pm0.3$ mag. If only sources in the $HF$ sub-sample are considered similar results hold. These results assure us that we are not introducing biases in clearing the frames from fitted sources.
\begin{table}
	\centering
	\caption{First, second and third quartiles of the distribution of the differences between properties derived without and with subtracting the source in the science frame after each fit.} 
	\label{tab:test_subtraction}
    \begin{tabular}{lrr}
    \hline
    \ & \multicolumn{2}{c}{NO SUB $-$ WITH SUB} \\
    Filter & \multicolumn{1}{c}{$25\ |\ 50\ |\ 75\%$}  & \multicolumn{1}{c}{$25\ |\ 50\ |\ 75\%$ $HF$} \\
    \hline
    $\rm R_c$ 	& $-0.21\ |\ \textbf{-0.02}\ |\ 0.21 $ 	& $-0.18\ |\ \textbf{-0.02}\ |\ 0.44 $ \\
    F140LP 		& $-0.34\ |\ \textbf{0.04}\ |\ 0.28 $ 	& $-0.36\ |\ \textbf{0.03}\ |\ 0.23 $ \\
    F150LP 		& $-0.17\ |\ \textbf{0.05}\ |\ 0.34 $ 	& $-0.36\ |\ \textbf{0.02}\ |\ 0.22 $ \\
    F336W 		& $-0.25\ |\ \textbf{0.04}\ |\ 0.22 $ 	& $-0.30\ |\ \textbf{0.03}\ |\ 0.21 $ \\
    F438W 		& $-0.20\ |\ \textbf{0.04}\ |\ 0.26 $ 	& $-0.38\ |\ \textbf{0.04}\ |\ 0.21 $ \\
    F775W 		& $-0.20\ |\ \textbf{0.04}\ |\ 0.26 $ 	& $-0.33\ |\ \textbf{0.04}\ |\ 0.23 $ \\
    \hline
    \end{tabular}
\end{table}

\subsection{Completeness limits}
\label{sec:test_completeness}
We tested the photometric completeness of the clumps in LARS galaxies filter by filter by inserting synthetic clumps of known properties (sizes and fluxes) in the LARS scientific frames and fitting them with the same photometric code used for the real clumps. 
For each filter in each galaxy, 1000 synthetic clumps were inserted at random positions inside the area of the galaxy (as defined in Section~\ref{sec:clumpiness}), with magnitudes uniformly sampled in the range $20-28$ mag and \r\ values sampled from a log-uniform distribution with boundaries $\rm r_c=0.3-3.5$ px. Sources were inserted and analysed one by one, in order to avoid biasing the results by artificially increasing the crowding. We point out that the source was inserted on top of the science frame and analysed without subtracting possibly brighter sources before the analysis, as was done in the pipeline for the real clumps. In this way the completeness we retrieve is an upper limit, as we know that sources at low luminosities benefit from the subtraction of nearby brighter sources.
We show the results of the completeness analysis in Fig.~\ref{fig:completeness}. We divide the sample of synthetic sources in bins of 1 mag width and consider as correctly recovered only the sources that satisfy the following conditions: 
\begin{equation}
\begin{split}
& \rm |coords_{in}-coords_{out}| < 1.5\ px \\
& \rm |mag_{in} - mag_{out}| < 0.5\ mag \\
& \rm |r_{c,in} - r_{c,out}| < 0.5\ px 
\end{split}
\end{equation}
We find completeness values that vary strongly from galaxy to galaxy, despite similar sets of observations, suggesting that completeness is mainly dependent on properties such as clump crowding and different contribution of galaxy diffuse emission.
We report in Tab.~\ref{tab:completeness} for each galaxy and filter the deepest magnitude above a $90\%$ completeness. 
\begin{table}
\centering
\caption{Completeness limits for all combinations of filters and galaxies, assuming a $90\%$ completeness limit. Data have been binned in 1 mag width bins to derived the listed values (see Fig~\ref{fig:completeness}). The completeness values for compact sources with $\rm r_c<2.0$ px are given within parentheses.} 
\label{tab:completeness}
\begin{tabular}{cccccc}
\hline
Name & F140LP & F150LP & $U-$band & $B-$band & $I-$band \\
\ & mag & mag & mag & mag & mag \\
\hline
L01 & 22 (23) & 22 (23) & 22 (24) & 22 (23) & 22 (23) \\
L02 & 23 (24) & 23 (24) & 23 (25) & 23 (25) & 23 (24) \\
L03 & 24 (25) & 23 (25) & 24 (25) & 23 (24) & 23 (24) \\
L04 & 24 (24) & 23 (24) & 24 (23) & 24 (24) & 23 (24) \\
L05 & 23 (23) & 21 (23) & 23 (24) & 22 (23) & 23 (23) \\
L06 & 24 (25) & 24 (25) & 24 (25) & 24 (25) & 24 (24) \\
L07 & 22 (23) & 22 (23) & 23 (24) & 22 (23) & 22 (23) \\
L08 & 24 (24) & 24 (24) & 24 (24) & 23 (24) & 22 (23) \\
L09 & 23 (23) & 23 (23) & 23 (24) & 22 (23) & 22 (23) \\
L10 & 24 (25) & 23 (25) & 24 (24) & 23 (25) & 23 (24) \\
L11 & 23 (24) & 22 (24) & 23 (24) & 23 (24) & 22 (23) \\
L12 & 24 (25) & 24 (26) & 24 (25) & 24 (25) & 23 (24) \\
L13 & $-$  & 24 (25) & 24 (25) & 24 (25) & 22 (23) \\
L14 & $-$  & 24 (25) & 24 (25) & 24 (25) & 23 (24) \\
\hline
\end{tabular}
\end{table}

We know that the completeness of the clumps is related to their surface brightness more than to the magnitude itself and for this reason we also plot the completeness curves considering only sources with $\rm r_c<2.0$ px. With this selection the completeness of the sample becomes deeper, typically by $\sim1$ mag.
\begin{figure*}
\centering
\subfigure{\includegraphics[width=\columnwidth]{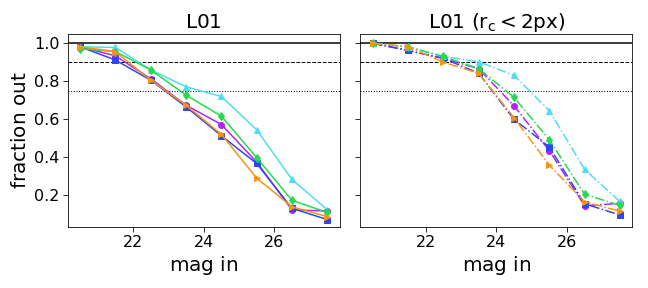}}
\subfigure{\includegraphics[width=\columnwidth]{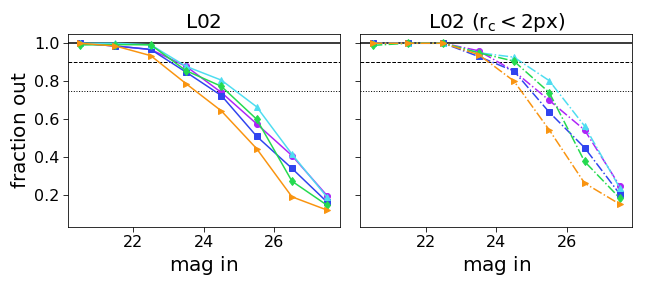}}
\subfigure{\includegraphics[width=\columnwidth]{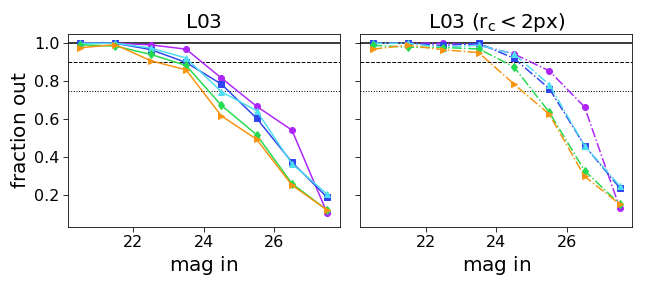}}
\subfigure{\includegraphics[width=\columnwidth]{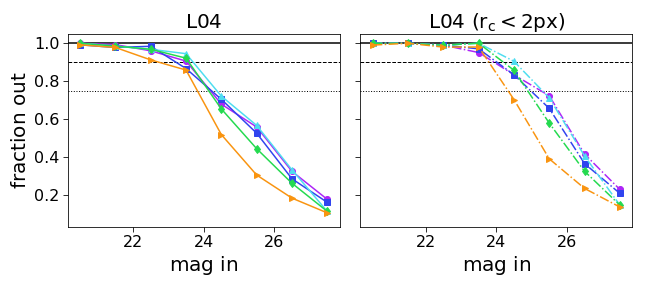}}
\subfigure{\includegraphics[width=\columnwidth]{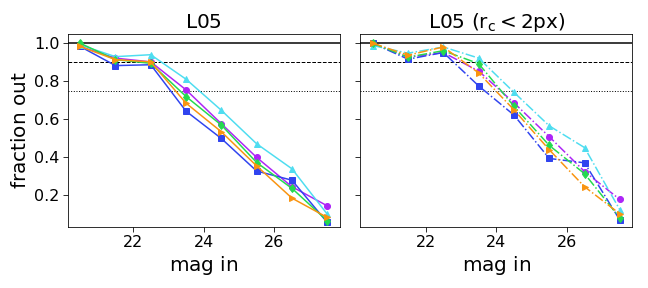}}
\subfigure{\includegraphics[width=\columnwidth]{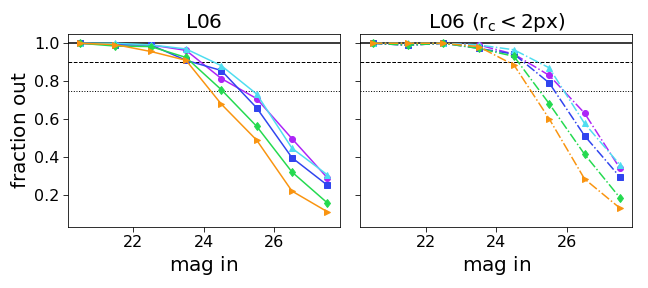}}
\subfigure{\includegraphics[width=\columnwidth]{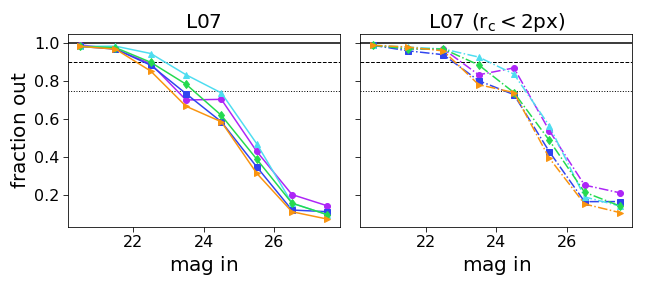}}
\subfigure{\includegraphics[width=\columnwidth]{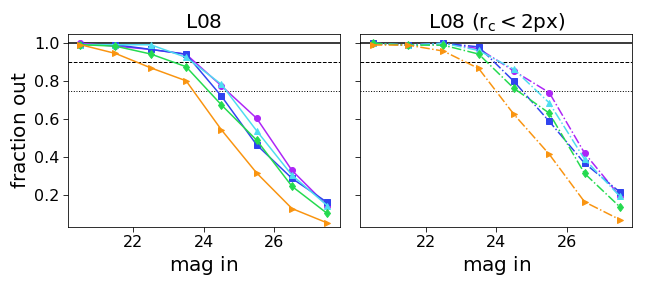}}
\subfigure{\includegraphics[width=\columnwidth]{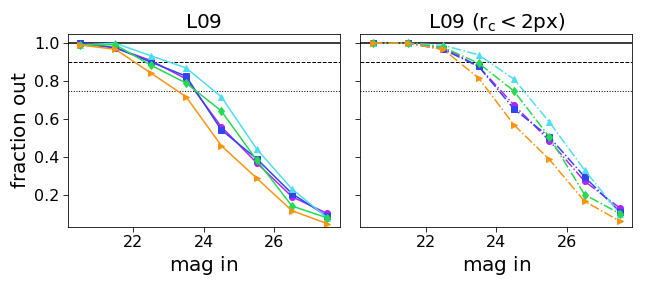}}
\subfigure{\includegraphics[width=\columnwidth]{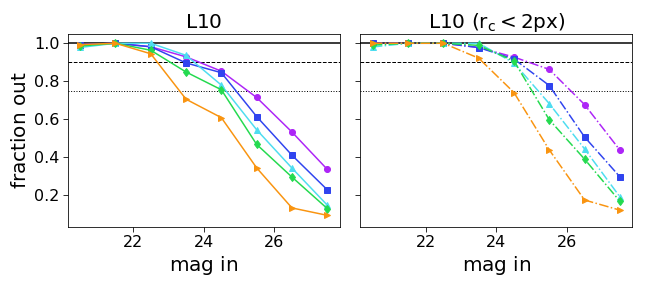}}
\caption{Completeness curves, plotted as the fraction of recovered sources versus the input magnitude. For each galaxy the left panel takes into account all the input sources, while the right one only the sources with input $\rm r_c<2$ px. Each filter is plotted as a separate curve, using the same colors of Fig.~\ref{fig:test_rcmin}, namely purple circles for F140LP, blue squares for F150LP, cyan triangles for the $U$ band, green diamonds for the $B$ band and orange triangles for the $i$ band. The horizontal lines mark completeness limits at 1 (solid), 0.9 (dashed) and 0.75 (dotted).}
\label{fig:completeness}
\end{figure*}
\begin{figure*}\ContinuedFloat
\centering
\subfigure{\includegraphics[width=\columnwidth]
{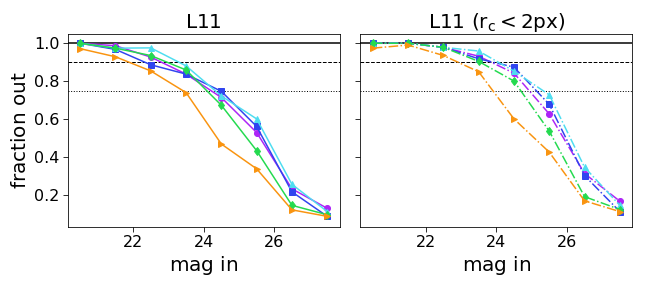}}
\subfigure{\includegraphics[width=\columnwidth]{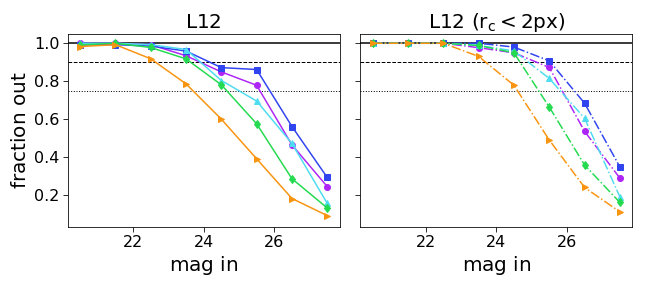}}
\subfigure{\includegraphics[width=\columnwidth]{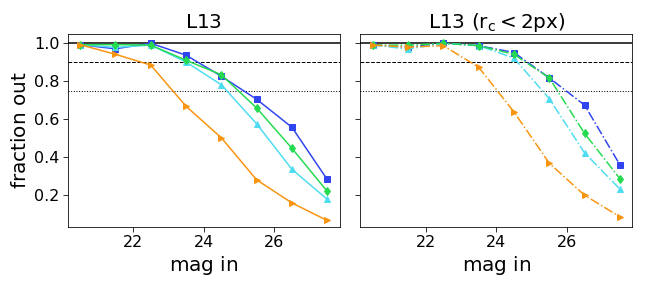}}
\subfigure{\includegraphics[width=\columnwidth]{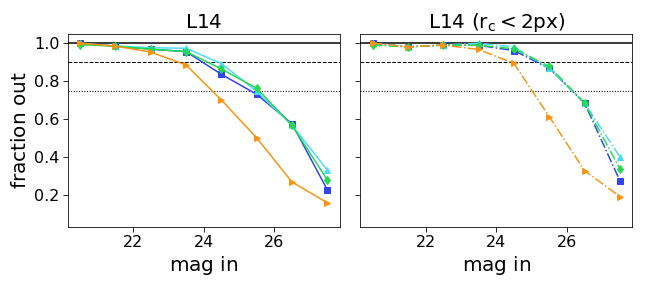}}
\caption{Continued.}
\end{figure*}

\section{Additional studies of the clump SFR}
\subsection{Clump SFR from de-reddened UV luminosity}
\label{sec:a2}
\begin{figure*}
\centering
\subfigure[]{\includegraphics[width=0.49\textwidth]{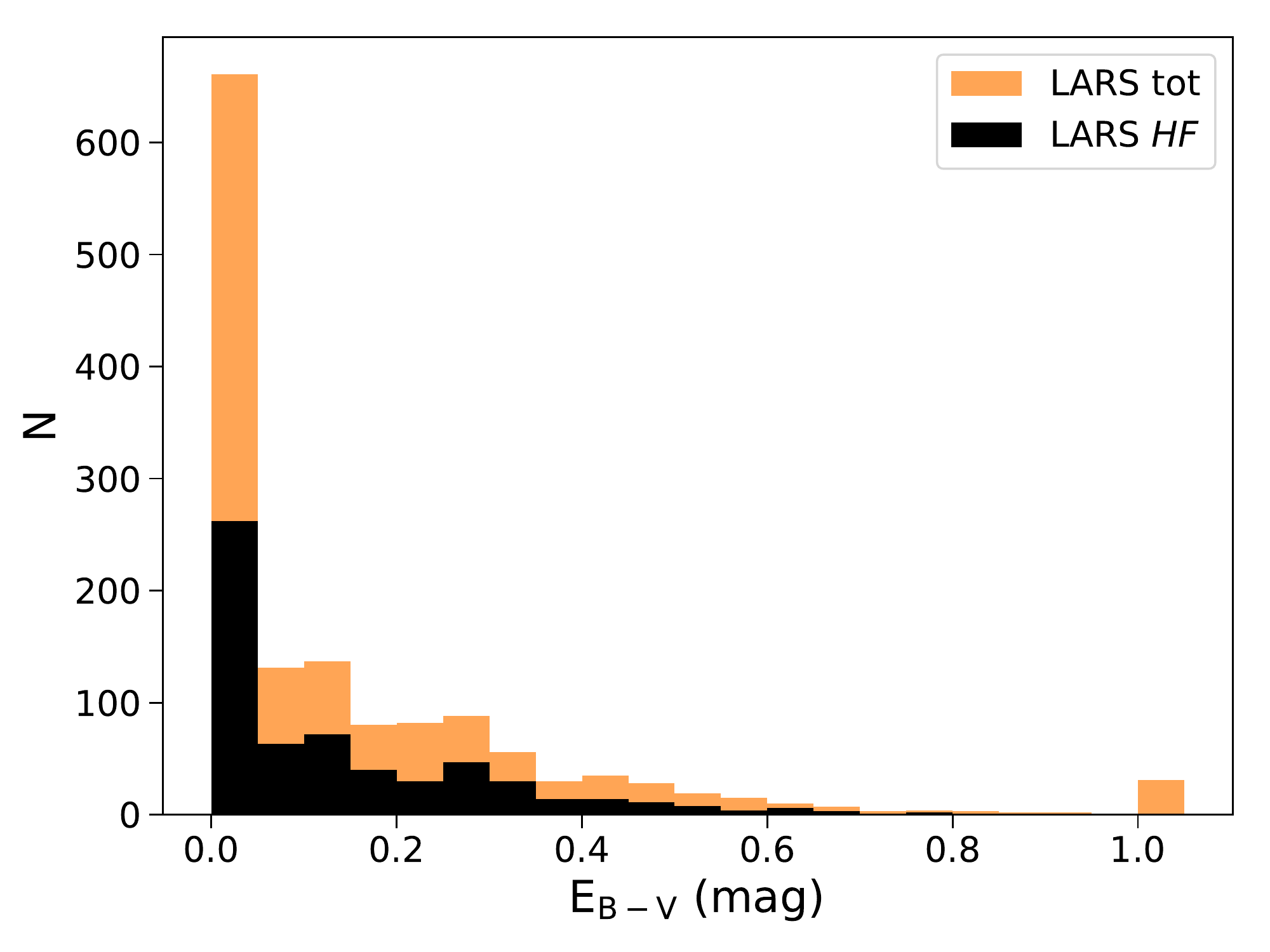}}
\subfigure[]{\includegraphics[width=0.49\textwidth]{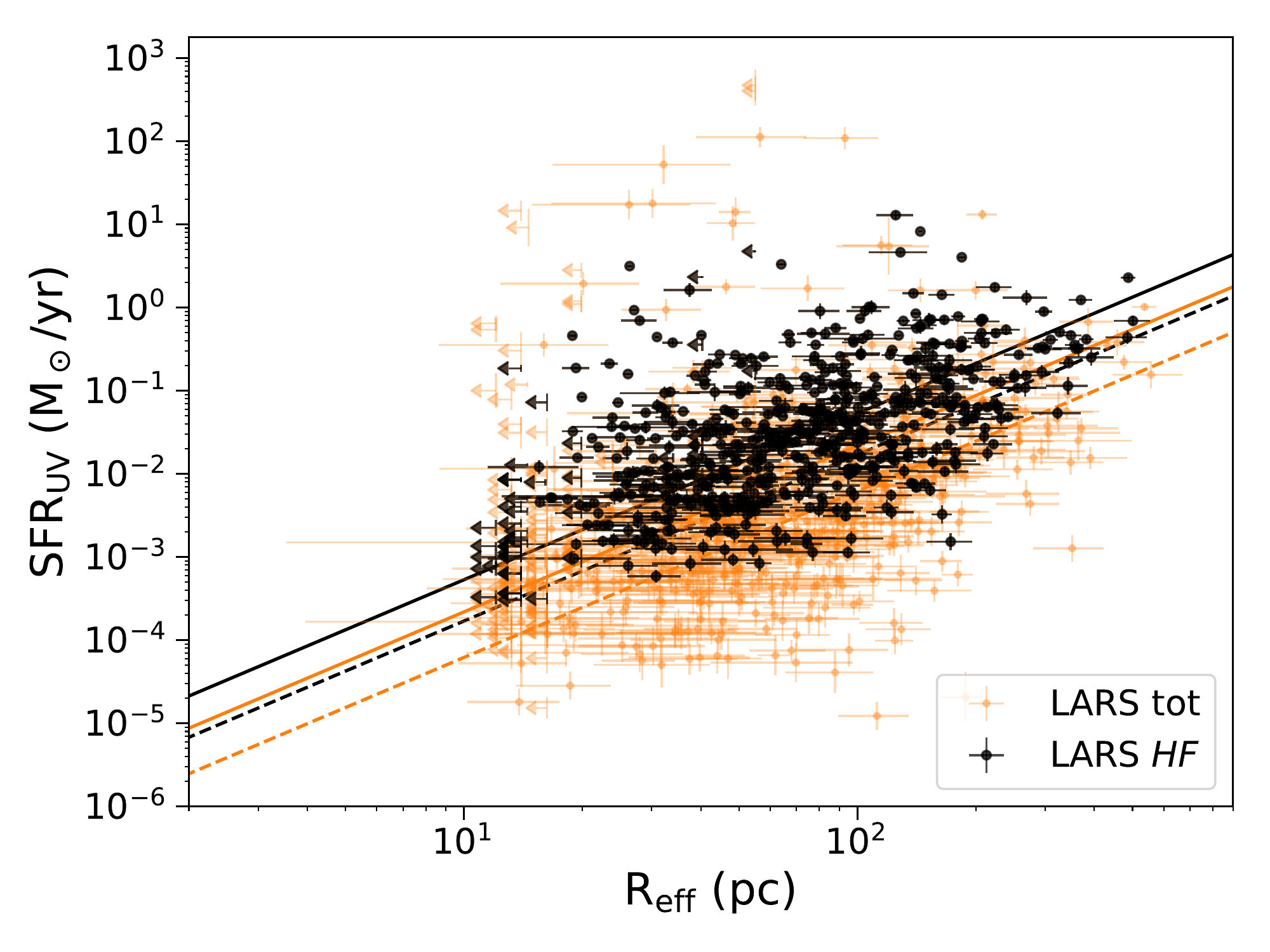}}
\caption{Panel (a): distribution of $\rm E_{B-V}$ values for the clumps in LARS, derived in Messa et al, (in prep.). Panel (b): updated version of Fig.~\ref{fig:size_lum}(a), using SFR values derived from $UV$ luminosities corrected for stellar extinction. The dashed lines are the old median values of \sigmasfr, while the solid lines are the new values.}
\label{fig:size_lum_dered}
\end{figure*}

in Section~\ref{sec:surface_brightness} we analysed the SFR of clumps derived from $UV$ luminosities, with the simplifying assuming that clumps have no extinction. In this appendix we repeat the analyses of Section~\ref{sec:surface_brightness}, de-reddening the $UV$ luminosity of each clump by its internal extinction, using the $\rm E_{B-V}$ values derived in Messa et al, (in prep). 
The luminosity is then converted into a SFR using the same \citet{kennicutt2012} relation used in Section~\ref{sec:surface_brightness}.
The distribution of the E(B-V) values for all LARS clumps is shown in panel a of Fig.~\ref{fig:size_lum_dered}. The distribution is peaked on zero extinction. 

Panel b of Fig.~\ref{fig:size_lum_dered} show the updated clumps' SFR-size relation, with the de-reddened SFR values.
The median SFR density of clumps increases $\sim3$ times, going from $\rm 0.20\ M_\odot/yr/kpc^2$ to $\rm 0.70\ M_\odot/yr/kpc^2$ for the total sample and from $\rm 0.54\ M_\odot/yr/kpc^2$ to $\rm 1.70\ M_\odot/yr/kpc^2$ for the $HF$ sub-sample, as can be appreciated in the figure. Therefore, even if the majority of clumps are weakly extincted, the effect on the clumps' \sigmasfr\ is significant.

\subsection{Fit of the clumps $\rm SFR-R_{eff}$ relation}
\label{sec:a22}
\begin{figure}
\centering
\includegraphics[width=\columnwidth]{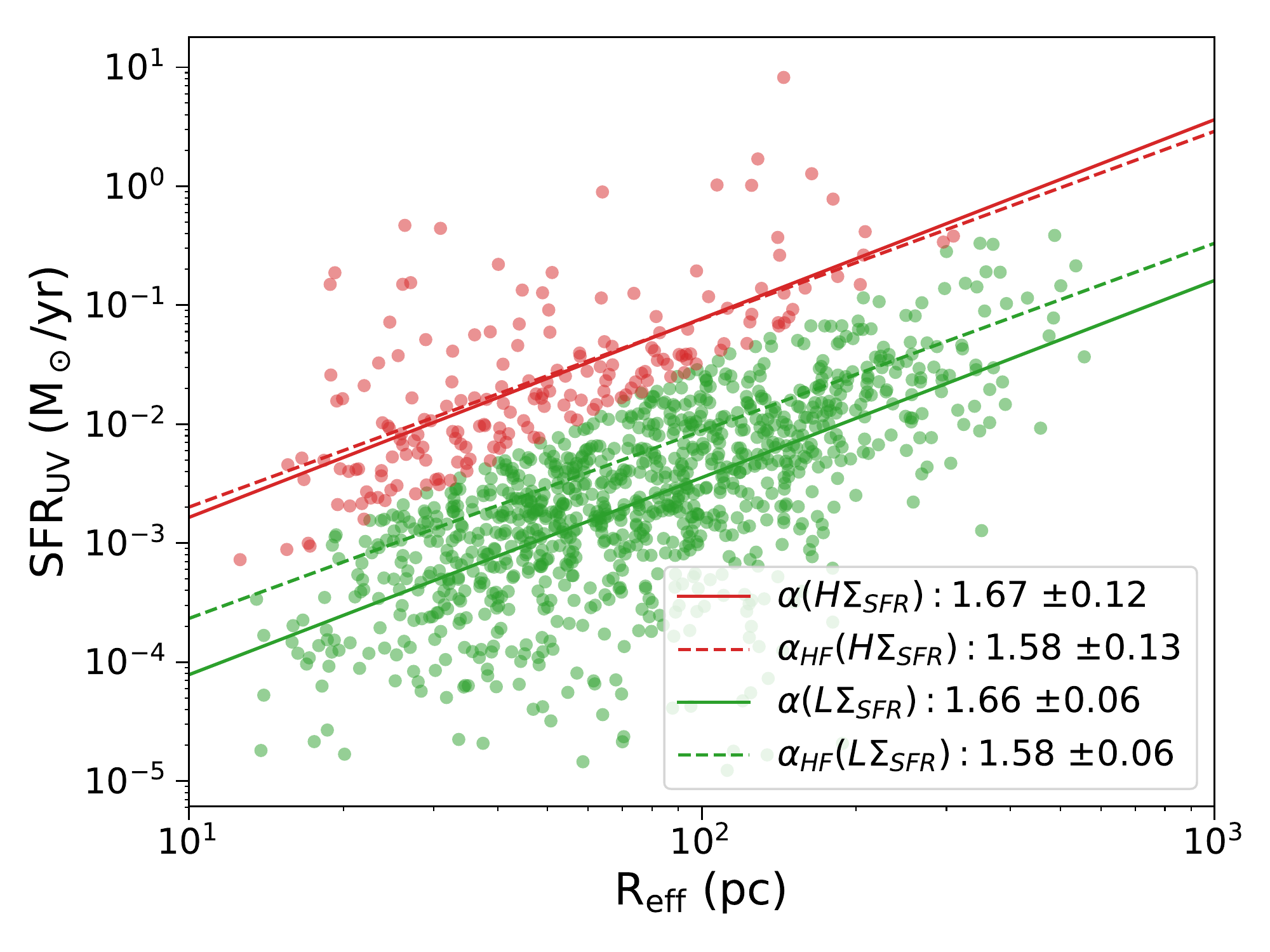}
\caption{SFR-size relation for the clumps in the LARS galaxies. Clumps with $\Sigma_{\textrm{SFR}}>1\ \textrm{M}_\odot/\textrm{yr}/\textrm{kpc}^2$ are colored in red. The best fit lines for the relation $\textrm{SFR}\propto \textrm{R}_{\textrm{eff}}^\gamma$ are over-plotted for the total sample (solid lines) and for the HF sub-sample (dashed lines).}
\label{fig:sfr_size}
\end{figure}

We fitted the clumps $\rm SFR-R_{eff}$ relation using the functional form $\textrm{L}_{\textrm{UV}}\propto \textrm{R}_{\textrm{eff}}^{\gamma}$. As described in the text we assumed the \citet{kennicutt2012} relation $\rm \log(SFR)\ [M_\odot/yr] = \log(L_{UV})\ [erg/s]- 43.35$ for the conversion between SFR and $UV$ luminosity. We assume for this fit no internal extinction for the clumps and we use a simple least-squares fitting method. We compute the fit both considering the total sample and the HF sub-sample. In both cases, the sample of clumps is divided in two according to their $\Sigma_{\textrm{SFR}}$ using $\rm 1\ M_\odot/yr/kpc^2$ as separating value. Results of the fit are given in Fig.~\ref{fig:sfr_size}. We do not find a considerable difference between the different sub-samples.

\subsection{Median SFR of clumps in sub-samples}
\label{sec:a23}
We divided the LARS galaxies in sub-samples in function of their properties and in particular focusing on their SFR surface density, $v_s/\sigma_0$ and stellar mass. For each property we dived the sample in two, using as separating values \sigmasfr=$\rm 0.22\ M_\odot/yr/kpc^2$, $v_s/\sigma_0=1$ and $\rm M_*=2\times10^{10}\ M_\odot$. The median clump \sigmasfr\ of each sub-sample is reported in Tab.~\ref{tab:median_sigmasfr}.
\begin{table}
\centering
\caption{Median \sigmasfr\ of the clumps. The columns are: (1) Name of the sub-sample, (2) ID of the galaxies included in the sub-sample, (3) median considering the clumps of the total sample, (4) median considering only the clumps of the HF sub-sample.} 
\label{tab:median_sigmasfr}
\begin{tabular}{llcc}
\hline
\multicolumn{1}{c}{Sample} & \multicolumn{1}{c}{Included} & \multicolumn{1}{c}{median(\sigmasfr)} & \multicolumn{1}{c}{median(\sigmasfr$\rm _{,HF}$)} \\
\multicolumn{1}{c}{} & \multicolumn{1}{c}{} & \multicolumn{1}{c}{($\rm M_\odot/yr/kpc^2$)} & \multicolumn{1}{c}{($\rm M_\odot/yr/kpc^2$)} \\
\multicolumn{1}{c}{(1)} & \multicolumn{1}{c}{(2)} & \multicolumn{1}{c}{(3)} & \multicolumn{1}{c}{(4)} \\
\hline
$\rm low\ \Sigma_{SFR}$     & 2,3,4,6,8,9   & $0.16$   & $0.45$    \\
$\rm high\ \Sigma_{SFR}$    & 1,5,7         & $0.60$   & $1.00$    \\
\hline
low $v_s/\sigma_0$          & 2,5,7         & $0.31$   & $1.12$  \\
high $v_s/\sigma_0$         & 1,3,4,6,8,9   & $0.19$   & $0.48$  \\
\hline
low $\rm M_*$               & 1,2,5,6,7     & $0.34$   & $1.00$  \\
high $\rm M_*$              & 3,4,8,9       & $0.18$   & $0.43$  \\
\hline
\end{tabular}
\end{table}



\bsp	
\label{lastpage}
\end{document}